\documentclass[journal]{IEEEtran}

\ifCLASSINFOpdf
   \usepackage[pdftex]{graphicx}
\else
\fi

\makeatletter
\newcommand*{\addFileDependency}[1]{
  \typeout{(#1)}
  \@addtofilelist{#1}
  \IfFileExists{#1}{}{\typeout{No file #1.}}
}
\makeatother

\usepackage{etoolbox}
\makeatletter
\patchcmd{\@makecaption}
  {\scshape}
  {}
  {}
  {}
\makeatletter
\patchcmd{\@makecaption}
  {\\}
  {.\ }
  {}
  {}
\makeatother

\usepackage{xr-hyper}

\usepackage{amsmath}
\usepackage{amsfonts}
\usepackage{amsthm}
\newtheorem{definition}{Definition}

\usepackage{hyperref}
\usepackage{multirow}
\usepackage[table,xcdraw]{xcolor}
\usepackage{float}
\usepackage{todonotes}
\usepackage{paralist}
\usepackage{footmisc}

\begin{document}

\title{The Portiloop: a deep learning-based open science tool for closed-loop brain stimulation}

\newcommand{\comment}[1]{}
\newcommand \TODO [1] {{\color{red}(#1)}}
\newcommand{\etal}{\textit{et al}.~}
\newcommand{\ie}{\textit{i}.\textit{e}., }
\newcommand{\eg}{\textit{e}.\textit{g}., }
\newcommand{\wrt}{w.r.t. }


\author{\IEEEauthorblockN{Nicolas Valenchon\IEEEauthorrefmark{1},
Yann Bouteiller\IEEEauthorrefmark{1},
Hugo R. Jourde\IEEEauthorrefmark{2},
Xavier L'Heureux\IEEEauthorrefmark{1},
Milo Sobral\IEEEauthorrefmark{1},
Emily B.J. Coffey\IEEEauthorrefmark{2} and
Giovanni Beltrame\IEEEauthorrefmark{1}
}

\IEEEauthorblockA{\IEEEauthorrefmark{1}MISTLab - Polytechnique, University of Montreal, Montreal, Quebec, Canada}

\IEEEauthorblockA{\IEEEauthorrefmark{2}CL:ASP, Concordia University, Montreal, Quebec, Canada}
}

\maketitle

\begin{abstract}

Closed-loop brain stimulation refers to capturing neurophysiological measures such as electroencephalography (EEG), quickly identifying neural events of interest, and producing auditory, magnetic or electrical stimulation so as to interact with brain processes precisely. It is a promising new method for fundamental neuroscience and perhaps for clinical applications such as restoring degraded memory function; however, existing tools are expensive, cumbersome, and offer limited experimental flexibility. In this article, we propose the Portiloop, a deep learning-based, portable and low-cost closed-loop stimulation system able to target specific brain oscillations. We first document open-hardware implementations that can be constructed from commercially available components. We also provide a fast, lightweight neural network model and an exploration algorithm that automatically optimizes the model hyperparameters to the desired brain oscillation. Finally, we validate the technology on a challenging test case of real-time sleep spindle detection, with results comparable to off-line expert performance on the Massive Online Data Annotation spindle dataset (MODA; group consensus). Software and plans are available to the community as an open science initiative to encourage further development and advance closed-loop neuroscience research\footnote{\label{foot1}\url{https://github.com/Portiloop}}.

\end{abstract}

\IEEEpeerreviewmaketitle

\section{Introduction}
Electrical activity within the brain forms the basis of perception, thought and
behaviour. This activity tends to be oscillatory in nature, as reciprocal
connections within and between brain regions form functional circuits for
processing and communicating information. Changes in electrical fields caused by
synchronously firing populations of neurons can be measured on the scalp using a
technique known as electroencephalography (EEG). Correlational studies have been
performed for nearly a century that attempt to link specific patterns and
frequency bands in EEG to cognitive functions or brain states. These approaches
are informative for many types of research questions and have increased our
understanding of brain processes, but they are unable to establish causal
relationships. The ability to interact with brain oscillations in a
precisely-timed fashion to enhance or inhibit endogenous processes - using
sensory~\cite{ngo2013auditory,ha2016multimodal,
  vonluhmann2017neural,kosmyna2019attentivu},
electrical~\cite{zarubin2018realtime} or
magnetic~\cite{shirinpour2020experimental} stimulation - allows for their
functional roles to be determined~\cite{zrenner2016closed}, and potentially for
restoration of processes deteriorated by aging or
pathology~\cite{vassileva2018neocortical}. These so-called \emph{closed-loop}
stimulation approaches thus hold great promise for neuroscience.

One of the closed-loop research areas that has progressed the fastest using
non-invasive neurophysiological recordings (i.e., EEG) and brain stimulation
techniques is studying memory consolidation processes in
sleep~\cite{ngo2013auditory, zotou2017realtime, ngo2019insights,
  chen2013realtime}. A first target has been slow oscillations (SOs: $0.5$ - $1.5$ Hz), which are
high amplitude waves that appear in non-rapid eye movement
(NREM) sleep and are known to be involved in memory
consolidation (i.e., the process by which recent learned experiences are transformed into long-term memory)~\cite{ngo2013auditory}. Using auditory stimulation to SO
up-states, when neural tissue is partly depolarized and more excitable,
Ngo~\etal enhanced the amplitude of SOs and reported an overnight improvement in
memory performance, a result that has now been replicated multiple times
(see~\cite{feher2021shaping, salfi2020boosting, harrington2021sounding} for
reviews). Closed-loop stimulation has also been used in the context of
preventing drowsiness~\cite{ha2016multimodal}, enhancing attention and
engagement~\cite{kosmyna2019attentivu}, and reducing central nervous system
damage after strokes~\cite{vonluhmann2017neural}.
There is great potential for these closed-loop stimulation techniques in
fundamental neuroscience, and potentially, for clinical applications
~\cite{zrenner2016closed, choi2020systematic}. However, progress is hampered by
the limited portability and flexibility of available systems, as well as by
their expense and by the complexity of their use.

The goal of our interdisciplinary collaboration between neuroscientists, data scientists and
computer engineers is to design, explore, and document the properties of a new,
complete closed-loop stimulation system (i.e., hardware and software), which we
call the \emph{Portiloop}. The Portiloop is a deep learning-based, portable,
battery-efficient and low-cost device that will enable the neuroscience
community to collect and process EEG data in real-time, detect patterns of
interest for fundamental research questions, and respond at low latency with
precisely-timed stimulation. We aim to accelerate fundamental research on
closed-loop stimulation in neuroscience by designing a highly functional device
and offering the code and plans freely to developers and scientists in the
research community.

The scope of this work encompasses both neuroscience and engineering aspects,
which may be of interest to audiences for different purposes. First, we describe
some general background concerning the use of closed-loop stimulation in
neuroscience and its potential, describe limitations in existing tools, and
introduce sleep spindles, a fast neural event that is observable in EEG, as a
challenging test case. Next, we discuss the real-time and portability design
constraints and the (hardware) architecture of our Portiloop implementation,
which is sufficiently powerful to allow us to run a neural network-based EEG
detection algorithm. The hardware is not commercially available in assembled
state, but it or a similar device may be constructed by readers with appropriate
technical training (plans and additional information are freely
available\footref{foot1}). Third, we describe a lightweight neural network
architecture that can run on inexpensive, modest hardware systems such as that
which we have proposed, and which can detect and react to physiological signals
in real time. Most importantly, we detail our design methodology and
optimization algorithm, so that the architecture can be adapted to other neural
events (e.g., theta or beta-band oscillations) or types of signal (e.g.,
functional near-infrared spectroscopy). This latter section and associated Supplementary Information will be of most
interest to readers with a data science background who may wish to implement,
use or modify the detection algorithm (all code is available\footref{foot1}). We then
present data from our case study EEG event, showing that the Portiloop
implementation can effectively detect sleep spindles in real time, and we
describe the performance with respect to detection threshold and time delay. The
latter sections may be most interesting for research users to understand the
performance of the system and select appropriate parameters for its use
detecting and stimulating brain oscillations. Finally, we discuss next steps and
future prospects for this technology.

The Portiloop is the first open-science device that is capable of closed-loop brain stimulation. Its most noteworthy contributions include: 
\begin{itemize}
\item Two open-hardware implementations that can be constructed from
  commercially available components (one using the Xilinx Pynq FPGA together
  with the HackEEG board and one using a custom board and a Google Coral
  System-on-Module neural accelerator)
\item A fast implementation of a recurrent neural network model that can be run on
  inexpensive hardware to detect events in physiological signals in real time
\item A design-space exploration algorithm that automatically optimizes the
  model hyperparameters to the neural event to be detected
\item A real-time spindle detector with accuracy comparable to offline analysis
  by experts
\end{itemize}

We hope that the Portiloop will increase research on closed-loop stimulation,
and continue to evolve and develop as a community-supported tool.

\section{General background}

\subsection{Limitations of current systems and design objectives}

Speed, expense, flexibility, and portability are important considerations for
designing a highly functional research-focused closed-loop system. The brain's
endogenous oscillations range from about 0.1 to 150 Hz. Depending on the
application and the neural event of interest, real-time constraints can vary
from a few ms~\cite{xu2014enhanced} to seconds~\cite{kosmyna2019attentivu}.
Currently available commercial systems that are capable of slow oscillation
closed-loop stimulation have difficulty accurately and precisely detecting and
stimulating faster, higher frequency neural events. Devices that are fast enough
and flexible enough for research purposes tend to be derived from high-end
systems used for real-time computing in other applications, e.g., in aerospace
and automotive industry~\cite{ngo2013auditory}, and are large and expensive.

Various portable devices have been developed to acquire and process EEG signals.
In McCrimmon et al. ~\cite{mccrimmon2017performance}, the authors developed a
low-cost device limited to acquisition. Other portable devices enable
closed-loop stimulation~\cite{ha2016multimodal, vonluhmann2017neural,
  kosmyna2019attentivu}, some also based on low-cost
hardware~\cite{zotou2017realtime}, but work with simple heuristics and are
generally not sufficiently powerful for complex signal detection algorithms such
as those based on deep learning. Our goal is to design a closed-loop system that
runs on inexpensive, portable hardware, yet is still sufficiently fast,
powerful, and flexible for cutting-edge research. Another element of
experimental flexibility that we incorporate into the design is the capability
to change the input and output signals. Thus, although our current focus is
EEG and auditory stimulation, an EEG trace could be exchanged for
another physiological signal like that derived from functional near infrared
spectroscopy, and detection output could be used to stimulate the brain more
forcefully using transcranial electrical or magnetic stimulation. By designing
the system flexibly such that it can be extended to detect and stimulate a
variety of brain oscillations, we can greatly expand its application, for
example to theta-band oscillations that are associated with working memory
capacity and task performance~\cite{albouy2017selective}, or sleep spindles. The
Portiloop is designed to be the first system to provide a portable, real-time
and deep learning-capable solution for multiple fundamental research
applications.

\subsection{Sleep spindles as a challenging test case}
\label{sec:casestudy}
Slow oscillations, which have been the main target for closed-loop auditory stimulation 
(CLAS) to date, are thought
to work in concert with other faster oscillations, called \emph{sleep spindles},
to reactivate recently learned memories and transfer them to long-term
memory~\cite{bergmann2017neuronal,rasch2013sleep}. Sleep spindles are transient
oscillations observed in both lighter and deeper non-rapid eye movement (NREM)
sleep (\ie sleep stages 2 and 3). Their role in memory consolidation is supported by
increases in spindle density following learning (\eg\cite{fogel2006learning}),
and the observation that age-related changes in sleep spindles are correlated
with differences in overnight performance gains (\eg\cite{lafortune2014sleep,
  fogel2017sleep}; see~\cite{fernandez2020sleep} for a review of spindle
mechanisms and functions).

If it were possible to influence spindles with sound, as it is to enhance slow
oscillations, researchers could explore their functional role in healthy adults
as well as characterize their involvement in cognitive aging, and even perhaps
restore degraded function. Particular challenges of spindle stimulation are that
each oscillatory cycle is only $\sim60$~ms long and the entire spindle is
between $0.5$ and $2.5$~s, leaving little time for traditional window-based
frequency analysis; there is considerable variability between the frequency,
amplitude, and duration of individuals' spindles, particularly in older
populations~\cite{peters2014age, purcell2017characterizing}; and even for
offline detection of spindles (which is an easier task than detecting spindles
online, as the entire spindle is available and can be used in detection),
agreement on spindle identification between experts themselves is limited
($\sim70$\%)~\cite{lacourse2019sleep, lacourse2020massive}. Real-time detection
of spindles is therefore a challenging test case for the Portiloop, and a
working online spindle detector would be of direct interest as a research tool.

\subsection{Offline sleep spindle detection for labeling and performance comparison}
\label{sec:offlinedetection}

Machine learning-based detection algorithms are powerful means of detecting
subtle signals in physiological data such as EEG, but they require large sets of
accurately labeled data for training and testing the algorithm's performance.
Once trained, the success of an algorithm on classifying previously unseen data
can be quantified using the f1-score, which is a widely used metric to quantify
an average of recall (i.e., success in detecting events) and precision (i.e.,
the proportion of detected events that are correct), see 
the Supplementary Information for equations.
The consistent detection and labeling of sleep spindles is a challenging task,
due to variability in their appearance and strength. Traditionally, spindles
have been visually identified by multiple experts, with f1-scores computed for
each scorer with respect to spindles identified by the consensus. One
commonly used dataset for creating and testing spindle detection
algorithms~\cite{kulkarni2019deep,tapia2020red} is the Montreal Archive of Sleep
Studies (MASS)~\cite{oreilly2014montreal}, in which the sleep spindle
annotations were provided by two experts. Projects using MASS for training
usually take spindles identified by either expert (i.e., a logical ``OR"
operation). However, the MASS annotations have a low inter-rater agreement
(f1-score = $0.54$~\cite{lacourse2020massive}), which makes this procedure statistically naive. The Massive Online Data
Annotation (MODA)~\cite{lacourse2020massive} project addressed this issue by
having 5 experts (on average) annotate spindles on a subset of data from MASS,
and rate their confidence, in each EEG segment. The experts had an inter-rater
f1-score of $0.72$ with respect to the final MODA labels. This score is
considerably better than the MASS equivalent, and the number of experts, the
scoring and the post-processing steps enable final labels of much higher
precision. We therefore adopt MODA as a basis for performance measurement,
bearing in mind that even MODA does not provide a true
answer about whether a spindle has occurred or not; only some degree of
consensus.

Several offline sleep spindle detectors have been developed and tested on
MODA~\cite{lacourse2019sleep,ferrarelli2007reduced,molle2002grouping,martin2013topography,wamsley2012reduced,ray2015expert,parekh2015detection}.
However, these generally use heuristics that compute Fourier transforms or
wavelet decomposition on large portions of the signal. For real-time detection
in online applications, spindles must be detected soon after their onset, if
stimulation is to arrive before the spindle ends and thus be capable of
influencing its evolution. Online real-time detectors therefore cannot take the
same approaches that have been successful for offline detection.

\subsection{Considerations for online sleep spindle detection}
\label{sec:onlineconsiderations}
Online detectors (\ie detectors that act during signal acquisition) face more
challenging conditions than offline detectors, due to the unavailability of
``future'' data points. For example, if we aim to detect and stimulate a spindle before it ends, the duration of the spindle is not yet
known by definition, yet it is one of the identifying criteria for spindles commonly used by experts. Some
existing heuristics filter the signal, compute power features and rely on
thresholds to perform detection; however, these approaches yield relatively poor
f1-scores~\cite{zotou2017realtime}.

Deep learning can also be leveraged to
perform online sleep spindle detection. This is done by first training an
artificial neural network offline through supervised learning to detect sleep
spindles, and then feeding the incoming signal to the trained detector. Several
such models have been trained in previous work~\cite{ kulkarni2019deep,
  tapia2020red, yasuhara2019study, tan2015sleep}. However, these works do not
consider hardware constraints that are central for our purpose: they use large
models that are often unable to run in real time even on high-end GPUs, which
makes them inapplicable in embedded systems. Moreover, they are usually trained
and tested on MASS~\cite{oreilly2014montreal} with an ``OR" operation performed
on the two experts' labels, which as discussed above is not a highly precise
target~\cite{lacourse2020massive}.

In this work, we design a Pareto-optimal neural architecture that performs best
on the MODA dataset~\cite{lacourse2020massive} while satisfying our hardware and
timing constraints. We validate our architecture against the state-of-the-art
SpindleNet~\cite{kulkarni2019deep}, initially used with the MASS dataset. When
both architectures are trained and tested on MODA, ours vastly outperforms the
baseline, on top of running in real time on embedded hardware.

\section{The Portiloop System}

\begin{figure*}
  \centering
  \includegraphics[trim=0 0 0 0, width=0.8\textwidth]{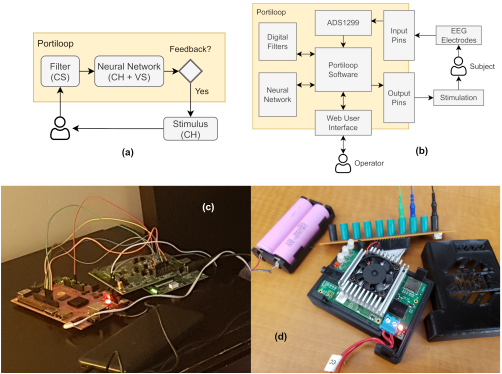}
   \caption{\textbf{Portiloop implementation.} (a) High-level view of the
     system: captured EEG data is first filtered and fed into a neural network.
     If a neural event is detected with sufficient confidence ($>$threshold), a
     decision is made to initiate brain stimulation. The types of delays
     introduced by each component are denoted in parenthesis: C/V denote
     Constant/Variable delays, and H/S denote Hardware/Software delays. (b) A
     detailed implementation scheme, and two possible implementations: (c) an
     FPGA prototype based on off-the-shelf components, and (d) a Coral-based
     implementation that uses a custom printed circuit board. Plans are
     available on our GitHub page. }
\label{fig:hardware}
\end{figure*}

A high-level description of the Portiloop system is provided in
Fig~\ref{fig:hardware}~(a), while a more detailed implementation scheme can be found in Fig~\ref{fig:hardware}~(b).
Fundamentally, it is made of an EEG frontend connected
to an embedded computer which reads the EEG signals, filters them, feeds the
filtered signal to an Artificial Neural Network (ANN) trained to detect specific
signals, and generates a stimulus when a target pattern is detected.

We propose two implementations of the Portiloop that can be replicated by
readers with the appropriate technical background:
\begin{itemize}
\item A version that can be fully built using off-the-shelf components based on a
  Xilinx Pynq FPGA board and an 8-channel HackEEG frontend (Fig~\ref{fig:hardware}~(c))
\item A custom printed circuit board (PCB) featuring an EEG frontend and a
  Google Coral neural accelerator (Fig~\ref{fig:hardware}~(d))
\end{itemize}
The detailed hardware implementation is out of the scope of this paper, but
readers can find all instructions and plans in our open-source
repository\footref{foot1}.

\begin{figure*}
  \centering
  \includegraphics[trim=0cm -0.5cm 0 0.0cm, width=0.8\textwidth]{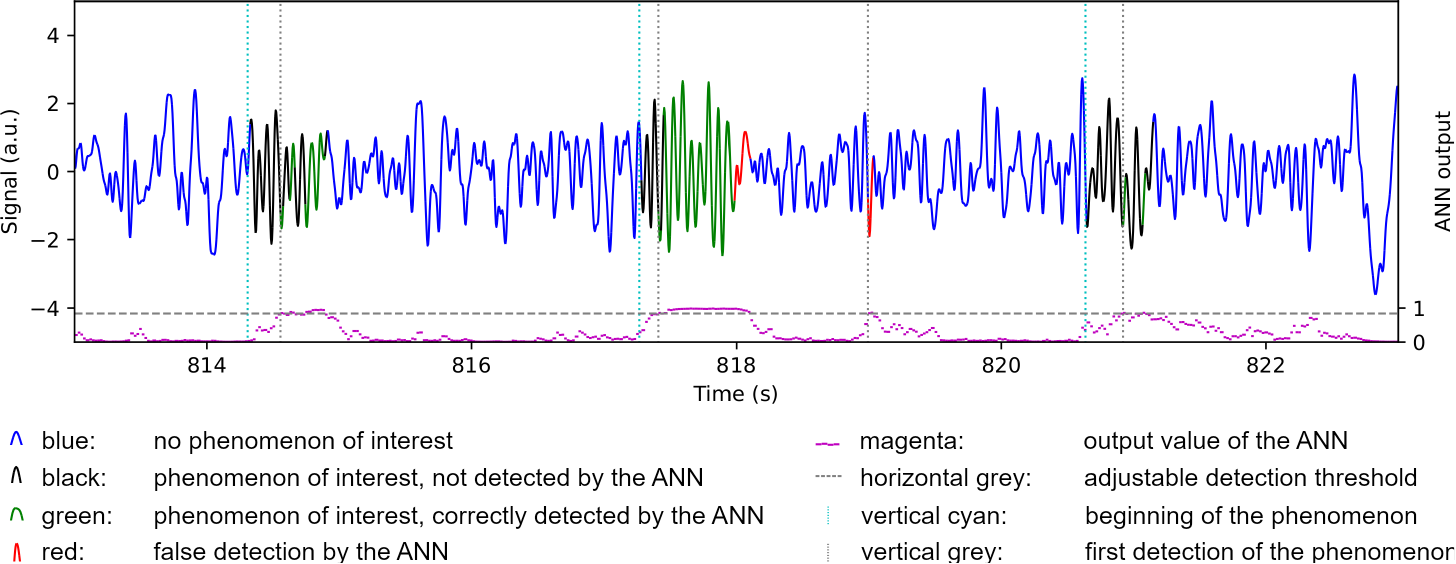}
  \caption{\textbf{Real-time stimulation example of the Portiloop on sleep 
  spindles.} The output of the ANN (likelihood between 0 and 1, magenta) is 
  displayed in the lower part of the Figure. When it crosses an adjustable 
  detection threshold (horizontal grey line, here set to 0.84), the Portiloop 
  sends a stimulus (vertical grey line). The optimal target for this stimulus 
  is the beginning of the sleep spindle (vertical cyan line). Thus, the 
  variable software delay is visible here between the vertical cyan and grey 
  lines.  The sections of the signal in red mark are sections wrongly detected
  as spindles (ANN false positives), and the areas in black those that are not, or not yet, detected as spindles but were identified as spindles by experts (ANN false negatives).
  }
\label{fig:software_delay}
\label{fig:stimu_84}
\end{figure*}

Since closed-loop stimulation requires very precise timing, the Portiloop needs
to detect target pattern as quickly as possible, and minimize the delay of
the output stimulus.
We identify two different sources of delay in the proposed system,
\emph{hardware} and \emph{software} delays. By \emph{hardware delays} we refer
to the time it takes to retrieve the signal from the electrodes, convert it to
digital, filter it, process it through the ANN, and send the resulting feedback
stimulation to the subject.
By \emph{software delays} we refer to time required for our system to collect
enough data to perform its functions. As an example, although the hardware
operations performed by signal filters are near-instantaneous, filtering
requires that a certain amount of data be collected before outputting a filtered value,
introducing a constant software delay in the output signal. This delay is a
trade-off related to the order of the filter. The higher the order of a filter,
the more efficient it is at removing undesirable frequencies, but also the
longer the software delay introduced in the signal by the filtering operation.
Similarly, an ANN may need to ``see'' a certain portion of a signal to
recognize it, introducing a (generally variable) delay on the output of the
classifier. An example of such delay is illustrated in
Fig~\ref{fig:software_delay}, where the trained ANN that we latter describe in Section~\ref{sec:results} takes a variable amount of time
before correctly detecting a transient pattern in EEG signal.
These hardware and software delays sum to a total delay that is the response time of the Portiloop system. They depend on the target
signal and put limits on the timing constraints of the application.

The Portiloop GitHub includes a software for recording and visualizing
the EEG signal on the device, as well as Python programming interface for the development of
extensions or new applications. The Portiloop can be accessed via
WiFi.
A web-based Graphical User Interface (GUI)
allows to configure the EEG channels, set up detection and stimulation, visualize the signals in real-time, record EEG, set up custom filters, and more.
The recording can be saved either in the internal memory (32GB) or an SD card in
EDF format, or streamed through the network using the Lab Streaming Layer
(LSL)\footnote{\url{https://github.com/sccn/labstreaminglayer}}, which
timestamps the data stream with microsecond accuracy.

\section{Neural Network implementation}

The Portiloop is primarily designed for EEG signals, i.e. time-series of data containing
oscillatory and transitory elements. In the realm of deep learning, a natural
way of processing such data is to use either 1D convolutions, recurrent units,
or a combination of both. The type of ANN architecture that we recommend is
inspired by SpindleNet~\cite{kulkarni2019deep}. In essence, a sliding window
over a few last data points is fed to a Convolutional Neural Network (CNN) whose
purpose is to extract relevant features (\eg frequencies) in this signal
fragment. Then, these extracted features are fed to a Recurrent Neural Network
(RNN) whose purpose is to keep track of the features extracted in past forward
passes (where a ``forward pass" is the action of computing an output from the ANN). Note that another family of architectures, called Transformers
\cite{vaswani2017attention}, is known for exhibiting good results with
this type of data when infinite compute is available for inference. However, Transformers are memory-less and not suitable for lightweight real-time applications, because they need to process the whole
signal at each forward pass. Conversely, RNNs are able to process one single
data point at each forward pass and keep track of the past in memory, which
makes them more applicable for the Portiloop.


The Portiloop has a limited amount of available memory, so as to ensure its
portability and low price. Therefore, large ANN architectures such as
SpindleNet \cite{kulkarni2019deep} are orders of magnitude too large to be
implemented in our device. To produce networks that are suitable to our device,
we rely on an automated optimization algorithm named ``Parallel Model-Based
Optimization'' (PMBO) that allows us to trade-off
accuracy and use of resources on our device (see Supplementary Information for details).

In addition, given the Portiloop's design constraints, we sought a lightweight
means of allowing our resource-restricted network to use as much signal history
as possible (as do larger neural networks). Time
dilation~\cite{chang2017dilated} is a technique that enables recurrent units
such as Gated Recurrent Units (GRUs) to look further back in time before
gradients vanish, at no computational cost. In the Supplementary Information, we propose a version of this
technique that allows us to virtually parallelize a single physical ANN into
several decoupled virtual models. Our approach enables shallow recurrent neural
networks to look further back in time by skipping the redundant information that
is inherent to the use of a sliding window as input, while still acting as fast
as possible.

\section{Case study: online sleep spindle detection}
We now turn our attention to a case study application of the Portiloop in
neuroscience - detecting sleep spindles shortly after they start so as to be
able to stimulate the brain during the spindle. The long-term goal of this
application is to further clarify the role of sleep spindles in learning and
memory, and to explore therapeutic interventions for memory decline (see Section
\ref{sec:casestudy}). As described in Section~\ref{sec:onlineconsiderations},
stimulating sleep spindles is a particularly challenging case study due to their
high frequency ($\sim$12 to 16~Hz) and rapid evolution ($<$2.5 s), and therefore
tight timing constraints, and thus serves as a demonstration of the technology's
capabilities.

To the best of our knowledge, the state-of-the-art in previous work regarding
online sleep spindles detection was SpindleNet~\cite{kulkarni2019deep}. This
architecture has too many parameters to be implemented on anything but the
largest graphics processing units. Moreover, it was trained and evaluated on the
MASS labels (i.e., a logic ``OR" on the annotations of two experts whose spindle
evaluation varies considerably). Since we do not have access to the SpindleNet
model, which is closed-source, we rebuilt the architecture described
in~\cite{kulkarni2019deep} and trained it on the more difficult MODA
dataset~\cite{lacourse2020massive} with the same pipeline that we used to train
our models, as a means of comparing the models' performance.

We draw inspiration from SpindleNet as a starting point for our ANN architecture
design. In particular, we train models based on the same idea of using
Convolutional Neural Networks (CNNs) followed by Recurrent Neural Networks
(RNNs), and we evaluate the relevance of the three different inputs used by
SpindleNet (namely, the raw signal, the signal envelope and the signal's power
features) in our setting. We then use our optimization algorithm (named PMBO)
along with the MODA dataset to derive a much smaller architecture, and provide a
quantitative comparison with the SpindleNet architecture on MODA. Since maximum
experimental flexibility is attained by being able to stimulate anytime during
the course of the spindle including with phase precision, we conduct a thorough
time analysis of the proposed system, and document possible trade-offs that a
researcher might use to maximize performance for a given experimental
application.

\subsection{Dataset and training}
\label{sec:training}
We use the MODA dataset (a subset of MASS), for training our ANN, since its
labels are considerably more reliable~\cite{lacourse2020massive}.
Ethical approval for use of the
dataset was obtained from the database's scientific committee and Concordia
University's Research Ethics Unit. This dataset is divided in two subsets. The
first one, called \emph{phase~1}, consists of 100 younger subjects, whereas the
second one, \emph{phase~2}, consists of 80 older subjects. The MODA dataset
provides two types of annotations (labels) on the signal: the first is the mean
score given by the group of experts for each data point; the second is a binary
classification of each data point as a spindle or non-spindle, defined by a
threshold on the aforementioned scores ($0.2$ for phase~1 and $0.35$ for
phase~2). Further post-processing steps were applied to obtain these binary
labels: spindles that were too short ($<0.3$~s) and too close ($<0.1$~s) to each
other were merged, then spindles that were too short ($<0.3$~s) or too long
($>2.5$~s) were relabeled as negative. Given this dataset, two types of ANNs are
possible: classifiers and regressors. These two types of ANNs differ only by the
labels and losses used to train them. Classifiers are trained on the binary
labels, by optimizing the binary cross entropy loss. They directly predict
whether the current signal is a spindle or not, according to the very specific
definition given by these binary labels (\ie taking into account the thresholds
and post-processing applied by MODA). Regressors are trained on the score
labels, by optimizing the mean square error loss. They predict the score given
by the experts (before the aforementioned post-processing steps), which allows
the user to select their own threshold for detection. Note that, in practice,
classifiers also enable the user to select their own threshold, although in a
less interpretable way. We experiment with both types of models. Finally, note
that MODA is a highly unbalanced dataset as only about 5~\% of the signal is
labeled as sleep spindles. During the course of this work, we tried different
ways of balancing training for classifiers and regressors. Interestingly, we
found that classifiers benefit highly from oversampling (\ie sampling 50~\% of
spindles and 50~\% of non-spindles from the dataset during training) whereas all
the balancing techniques we tried for regression (including oversampling, Label
Distribution Smoothing~\cite{yang2021delving} and a custom version of the
latter) actually hinder training.

\begin{figure}
  \centering
  \includegraphics[trim=0 0 0 0, width=1.0\linewidth]{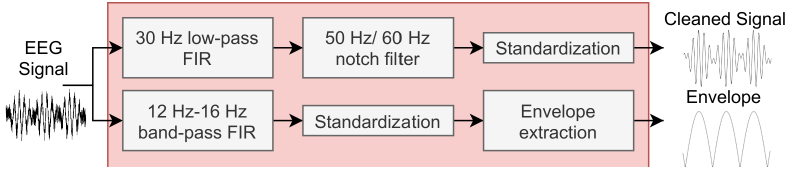}
  \caption{\textbf{Portiloop signal processing pipeline for extracting relevant inputs for the ANN.} The selected filters introduce an identical software delay of 40~ms in both branches. (Note that power features are computed offline only for the SpindleNet architecture and are not represented in this diagram.)}
\label{fig:datagen_pipeline}
\end{figure}

To evaluate against SpindleNet we compute the inputs used by this model: the
signal, the envelope of the signal, and a ``power feature
ratio''~\cite{kulkarni2019deep}. The latter compares frequencies between 2~Hz
and 8~Hz with frequencies between 9~Hz and 16~Hz from the Fourier transform over
the last 500~ms of signal. Computing this ratio is resource-intensive in the
context of the Portiloop system, and furthermore did not improve our models'
performance. Therefore, we compute this ratio offline for the sole purpose of
comparison with SpindleNet, and we do not use it in our model. We set the
sampling frequency to 500~Hz, which allows the Portiloop to log the raw signal
at a higher resolution, and then downsample to 250~Hz.
Fig~\ref{fig:datagen_pipeline} depicts the pipeline that computes the cleaned
signal and envelope.

We filter the EEG signal in the same frequency band as used in standard sleep
scoring (\ie $0.5$~Hz to 30~Hz)~\cite{iber2007aasm}.
An FIR filter of order 20 works reasonably well to remove frequencies above
30~Hz, but we observed persistent power line noise in unshielded home or office recording environments. To address this issue, we
apply a notch filter whose frequency depends on the geographical area (50~Hz in
Europe, 60~Hz in North America). For removing low frequencies, we rely on
online standardization through exponential moving average (formulas are provided in Supplementary Information).
We use a coefficient $\alpha_\mu=0.1$ for our running average, which attenuates
frequencies under 4~Hz. We use a smaller coefficient $\alpha_\sigma=0.001$ for
our running variance, meaning that our estimate of the standard deviation takes
a larger portion of the signal into account. We empirically found this choice of
$\alpha_\mu$ and $\alpha_\sigma$ to reveal EEG features of interest and yield
acceptable standardization, by visual inspection.

We apply a similar procedure to extract the envelope: first, we filter the
signal with a FIR band-pass between 12~Hz and 16~Hz. Then, we standardize with
$\alpha_\mu=\alpha_\sigma=0.001$, we square the signal, and we smooth the
result by computing its moving average, this time with $\alpha_\mu=0.01$.
We evaluate different types of ANN architectures, using either both or only one
of these preprocessed signals as input. Since FIR filters introduce software
delays, we have designed both branches of the pipeline so that they introduce
identical software delays to their respective outputs (\ie 40~ms at 250~Hz
sampling rate with FIR filters of order 20).

The output of the ANN tells whether the model considers the current
signal being a sleep spindle or not.  Some further processing is
necessary to ensure that we only send one stimulation per spindle.  As
seen in Fig~\ref{fig:software_delay}, the detection can be noisy
around the beginning or the end of a spindle, especially since we use
decoupled virtual parallel networks (see Supplementary Information).  
A stimulus is sent upon initial spindle detection. To avoid multiple stimuli of
the same spindle, the subsequent stimulation may only occur 400 ms following the
end of the spindle. If a spindle is detected again within this duration the
timer is reset, and we consider it as being part of the previous spindle.

\section{Validation and performance}
\label{sec:results}

We report results from a thorough quantitative and qualitative study of the system,
not only in terms of detection scores as generally seen in previous work (i.e, proportion
of data points correctly detected to be part of a spindle), but
also in terms of real-time stimulation performance. Note that all our
experiments are based on the MODA dataset rather than actual nights spent
wearing the Portiloop device, as we would not have ground truth labels for newly recorded
data. Further validation of the final device will require reproducing the experimental setting of MASS/MODA with the Portiloop (while participants are simultaneously wearing a research-grade
polysomnography system for comparison) and labelling
acquired data.

All results regarding online detection performance are summarized in
Table~\ref{tab:results}. This table shows the f1, precision and recall metrics
that statistically describe how efficient different models are at detecting
sleep spindles (on average over all data points). These metrics are provided
separately for phase~1, which groups younger subjects, for phase~2, which groups
older subjects, and for the whole cohort.

\begin{table*}
\centering
\caption{Quantitative results. Our different models and ablations are compared under ``Online Detection'' using the nomenclature ``mean (std)'', and superscripts for referencing rows in the text. In rows (4) and (5) we replace an input of our 2-input model by a copy of the other, in row (7) we remove time-dilation, in rows (8) and (9) we train our model only on phase 1 or phase 2 (i.e., young subjects or old subjects), and in row (10) we train a regressor to evaluate it as a classifier.}
\label{tab:results}
\resizebox{\textwidth}{!}{%
\begin{tabular}{c|c|c|
>{\columncolor[HTML]{E7E6E6}}c |c|c|
>{\columncolor[HTML]{E7E6E6}}c |c|c|
>{\columncolor[HTML]{E7E6E6}}c |}
\cline{2-10}
\multicolumn{1}{l|}{} & \multicolumn{3}{c|}{\cellcolor[HTML]{CCCCCC}\textbf{(a) Phase 1 (younger)}} & \multicolumn{3}{c|}{\cellcolor[HTML]{CCCCCC}\textbf{(b) Phase 2 (older)}} & \multicolumn{3}{c|}{\cellcolor[HTML]{CCCCCC}\textbf{(c) Whole Cohort}} \\ \cline{2-10} 
\multicolumn{1}{l|}{} & \cellcolor[HTML]{E5E5E5}\textit{\textbf{Recall}} & \cellcolor[HTML]{E5E5E5}\textit{\textbf{Precision}} & \cellcolor[HTML]{E5E5E5}\textit{\textbf{f1}} & \cellcolor[HTML]{E5E5E5}\textit{\textbf{Recall}} & \cellcolor[HTML]{E5E5E5}\textit{\textbf{Precision}} & \cellcolor[HTML]{E5E5E5}\textit{\textbf{f1}} & \cellcolor[HTML]{E5E5E5}\textit{\textbf{Recall}} & \cellcolor[HTML]{E5E5E5}\textit{\textbf{Precision}} & \cellcolor[HTML]{E5E5E5}\textit{\textbf{f1}} \\ \cline{2-10} 
 & \multicolumn{9}{c|}{\cellcolor[HTML]{C0C0C0}Experts} \\ \hline
\multicolumn{1}{|c|}{\cellcolor[HTML]{CCCCCC}\textit{Inter-rater agreement$^1$}} & 0.76 (0.16) & 0.81 (0.17) & \textbf{0.76 (0.1)} & 0.66 (0.19) & 0.74 (0.17) & \textbf{0.65 (0.12)} & 0.72 (0.18) & 0.78 (0.17) & \textbf{0.72 (0.12)} \\ \hline
 & \multicolumn{9}{c|}{\cellcolor[HTML]{C0C0C0}Offline Detection} \\ \hline
\multicolumn{1}{|c|}{\cellcolor[HTML]{CCCCCC}\textit{Ferrarelli\cite{ferrarelli2007reduced}}} & 0.19 & 0.83 & 0.31 & 0.16 & 0.87 & 0.27 & 0.18 & 0.85 & 0.29 \\ \hline
\multicolumn{1}{|c|}{\cellcolor[HTML]{CCCCCC}\textit{Mölle\cite{molle2002grouping}}} & 0.83 & 0.47 & 0.6 & 0.78 & 0.44 & 0.56 & 0.81 & 0.46 & 0.58 \\ \hline
\multicolumn{1}{|c|}{\cellcolor[HTML]{CCCCCC}\textit{Martin\cite{martin2013topography}}} & 0.61 & 0.64 & 0.62 & 0.58 & 0.56 & 0.57 & 0.6 & 0.6 & 0.6 \\ \hline
\multicolumn{1}{|c|}{\cellcolor[HTML]{CCCCCC}\textit{Wamsley\cite{wamsley2012reduced}}} & 0.57 & 0.69 & 0.63 & 0.56 & 0.62 & 0.59 & 0.57 & 0.66 & 0.61 \\ \hline
\multicolumn{1}{|c|}{\cellcolor[HTML]{CCCCCC}\textit{Lacourse\cite{lacourse2019sleep}}} & 0.75 & 0.73 & \textbf{0.74} & 0.7 & 0.69 & 0.7 & 0.73 & 0.71 & \textbf{0.72} \\ \hline
\multicolumn{1}{|c|}{\cellcolor[HTML]{CCCCCC}\textit{Ray\cite{ray2015expert}}} & 0.73 & 0.47 & 0.57 & 0.75 & 0.32 & 0.45 & 0.74 & 0.4 & 0.51 \\ \hline
\multicolumn{1}{|c|}{\cellcolor[HTML]{CCCCCC}\textit{Parekh\cite{parekh2015detection}}} & 0.85 & 0.61 & 0.71 & 0.74 & 0.68 & \textbf{0.71} & 0.8 & 0.65 & 0.71 \\ \hline
 & \multicolumn{9}{c|}{\cellcolor[HTML]{C0C0C0}Online Detection} \\ \hline
\multicolumn{1}{|c|}{\cellcolor[HTML]{CCCCCC}\textit{Based on SpindleNet\cite{kulkarni2019deep}$^2$}} & 0.92 (0.04) & 0.24 (0.07) & 0.38 (0.07) & 0.85 (0.06) & 0.19 (0.08) & 0.3 (0.1) & 0.89 (0.05) & 0.22 (0.07) & 0.35 (0.08) \\ \hline
\multicolumn{1}{|c|}{\cellcolor[HTML]{CCCCCC}\textbf{2-input$^3$}} & 0.68 (0.04) & 0.6 (0.06) & \textbf{0.64 (0.03)} & 0.52 (0.09) & 0.58 (0.04) & 0.54 (0.05) & 0.62 (0.06) & 0.6 (0.05) & \textbf{0.61 (0.03)} \\ \hline
\multicolumn{1}{|c|}{\cellcolor[HTML]{CCCCCC}\textbf{2-input ablation 1$^4$}} & 0.7 (0.09) & 0.47 (0.08) & 0.55 (0.04) & 0.56 (0.11) & 0.43 (0.09) & 0.47 (0.04) & 0.65 (0.1) & 0.46 (0.08) & 0.52 (0.04) \\ \hline
\multicolumn{1}{|c|}{\cellcolor[HTML]{CCCCCC}\textbf{2-input ablation 2$^5$}} & 0.72 (0.03) & 0.57 (0.06) & \textbf{0.64 (0.03)} & 0.57 (0.08) & 0.53 (0.04) & \textbf{0.55 (0.04)} & 0.67 (0.04) & 0.56 (0.05) & \textbf{0.61 (0.03)} \\ \hline
\multicolumn{1}{|c|}{\cellcolor[HTML]{CCCCCC}\textbf{1-input$^6$}} & 0.7 (0.04) & 0.59 (0.05) & \textbf{0.64 (0.03)} & 0.54 (0.09) & 0.58 (0.05) & \textbf{0.55 (0.05)} & 0.64 (0.05) & 0.59 (0.05) & \textbf{0.61 (0.03)} \\ \hline
\multicolumn{1}{|c|}{\cellcolor[HTML]{CCCCCC}\textbf{1-input ablation td$^7$}} & 0.47 (0.1) & 0.6 (0.09) & 0.51 (0.03) & 0.31 (0.12) & 0.59 (0.08) & 0.39 (0.08) & 0.41 (0.1) & 0.6 (0.09) & 0.47 (0.04) \\ \hline
\multicolumn{1}{|c|}{\cellcolor[HTML]{CCCCCC}\textbf{1-input trained on p1$^8$}} & 0.72 (0.05) & 0.56 (0.05) & 0.63 (0.03) & 0.57 (0.08) & 0.52 (0.07) & 0.54 (0.05) & 0.66 (0.07) & 0.55 (0.05) & 0.6 (0.03) \\ \hline
\multicolumn{1}{|c|}{\cellcolor[HTML]{CCCCCC}\textbf{1-input trained on p2$^9$}} & 0.75 (0.05) & 0.5 (0.05) & 0.6 (0.02) & 0.62 (0.09) & 0.45 (0.05) & 0.52 (0.03) & 0.7 (0.06) & 0.49 (0.04) & 0.57 (0.02) \\ \hline
\multicolumn{1}{|c|}{\cellcolor[HTML]{CCCCCC}\textbf{1-input regression$^{10}$}} & 0.62 (0.07) & 0.64 (0.06) & 0.63 (0.03) & 0.53 (0.06) & 0.55 (0.08) & 0.53 (0.04) & 0.58 (0.06) & 0.62 (0.06) & 0.6 (0.03) \\ \hline
\end{tabular}%
}
\end{table*}

As previously highlighted, sleep spindle detection is a difficult task and
experts themselves often do not agree when annotating these offline. This
disagreement is quantified by MODA~\cite{lacourse2020massive} and represented in
Table~\ref{tab:results}, row (1) for reference. The experts annotating the MODA
dataset had an average performance of $0.72$ on the whole cohort in term of the
f1-score of their individual annotations with respect to the final labels. They
are compared to other \emph{offline detection}, \ie when a virtually infinite
computational budget and the whole signal is available, including future data
points, presented under ``offline detection'' in Table~\ref{tab:results} (taken
from~\cite{lacourse2020massive}).
We instead perform \emph{online detection},
which has additional challenges:
\begin{inparaenum}[(a)]
    \item computation happens in real time;
    \item the future signal is not available.
\end{inparaenum}

The MODA dataset is relatively small ($\sim$24~h of annotated data) and
heterogeneous. This adds some difficulty for training and properly assessing the
performance of our models, because we choose to use only 10\% of subjects as our
validation set (for model selection), and another 10\% of subjects as our test
set (for final model evaluation).
Since the results would otherwise be dependent on the
assignment of subjects to the three sets, we evaluate our models through the
following procedure:
\begin{itemize}
\item we shuffle all subjects 10 times and compute a different
  training/validation/test split of the dataset each time (sets are thus made of
  separate subjects);
\item for each split, we use the training set to train 3 models, the validation
  set being used to estimate their f1-score. We select the best of these 3
  models by its best f1-score on the validation set. We then report the
  performance of this model in terms of its f1-score on the test set;
\item the above being repeated 10 times, we report the average test f1-score in
  Table~\ref{tab:results}, with the corresponding standard deviation being
  indicated in parenthesis.
\end{itemize}

As described previously, we use the SpindleNet~\cite{kulkarni2019deep}
architecture as a baseline for evaluating the performance of our own models.
Since SpindleNet is closed-source and trained on the MASS dataset, we retrain
its architecture from scratch with the same pipeline as used to train our other
classifiers. In particular, we balance training through oversampling (as opposed
to the data augmentation technique used by the authors of the original paper),
and we train and evaluate SpindleNet on the MODA dataset. The results of this
experiment are presented in Table~\ref{tab:results}, row (2). The baseline has a
high recall and a poor precision; in other words it it tends to incorrectly
label non-spindle events as spindles.

We first derive a lightweight ANN architecture by drawing inspiration from
SpindleNet. More precisely, we use our optimization algorithm PMBO to find a
Pareto-optimal architecture that uses both the cleaned signal and the envelope
as inputs. The resulting architecture is presented in the Supplementary
Information. We measure a total duration of 40~ms for each forward pass in this
model on the FPGA-based variant of the Portiloop. The detection performance of
this model, reported in Table~\ref{tab:results}, row (3), vastly outperforms the
baseline.

The idea of using the envelope along the raw signal as input to the ANN is drawn
from the baseline. Since the envelope is computed from the raw signal, it should
not contain any additional information that cannot be extracted by an ANN. To
evaluate the relevance of this particular input, we perform the following
ablation: to keep the same architecture (and thus the same model capacity), we
replace one of the two inputs by a copy of the other. In
Table~\ref{tab:results}, row (4) both inputs are the envelope, while in row (5)
both inputs are the cleaned signal. We find that the envelope input can be
removed: the model in which we replace the envelope with a copy of the cleaned
signal (5) has the same performance as the original model (3), and even performs
marginally better on phase~2.

\begin{figure}
  \centering
  \includegraphics[trim=0.5cm 0cm 1.5cm 0.8cm, clip, width=1.0\linewidth]{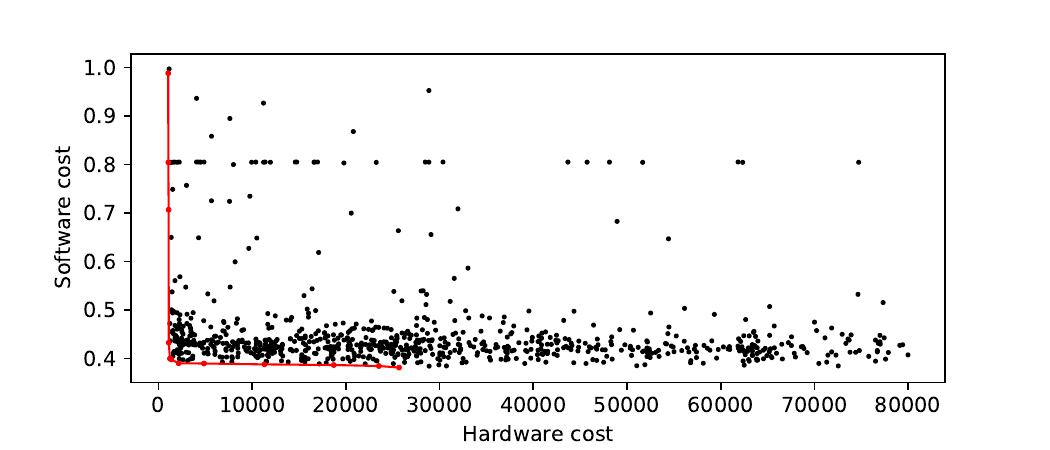}
  \caption{\textbf{Search space of the single-input architecture, found with PMBO.} The hardware
    cost is the number of trainable parameters in the neural architecture, and
    the software cost is $1-\text{f1-score of the fully-trained model}$. Black dots:
    non-Pareto-optimal models tested by the algorithm. Red dots: Pareto-optimal
    models found by the algorithm. Red line: Pareto front. The researcher would select a configuration from the Pareto front, which represents optimal trade-offs between both costs.}
\label{fig:pareto}
\end{figure}

Since we deem the use of the envelope input ineffective, we use PMBO one more
time to devise our final Pareto-optimal ANN architecture, now with only the
cleaned signal as input. For this matter, we run PMBO on 20 Tesla~V100 GPU
workers over a period of 24h. The detailed hyperparameters used in this
experiment are provided in Supplementary Information, and the results are
visualized in Fig~\ref{fig:pareto}, which shows all the explored architectures
according to their classification performance (software cost) and the use of
FPGA resources (hardware cost). The red line is the Pareto front, meaning the
set of configurations that are optimal for at least one of the two metrics: this
means all points that are \emph{not} on the Pareto front have at least one
corresponding configuration that is better in terms of both software and
hardware cost, and should therefore not be considered. We select the best model
in terms of software cost (i.e. the one with the highest classification
performance) irrespective of its hardware cost \ie the model corresponding to
the right-hand end of the Pareto front. This model is acceptable because it is
anyway rather small, with only 25.6k parameters. We measure the execution time
of this architecture to be 20~ms per forward pass on the Portiloop (vs. 40~ms
for the 2-input version).

\begin{figure}
  \centering
  \includegraphics[trim=0cm 0cm 0cm 0cm, clip, width=0.8\linewidth]{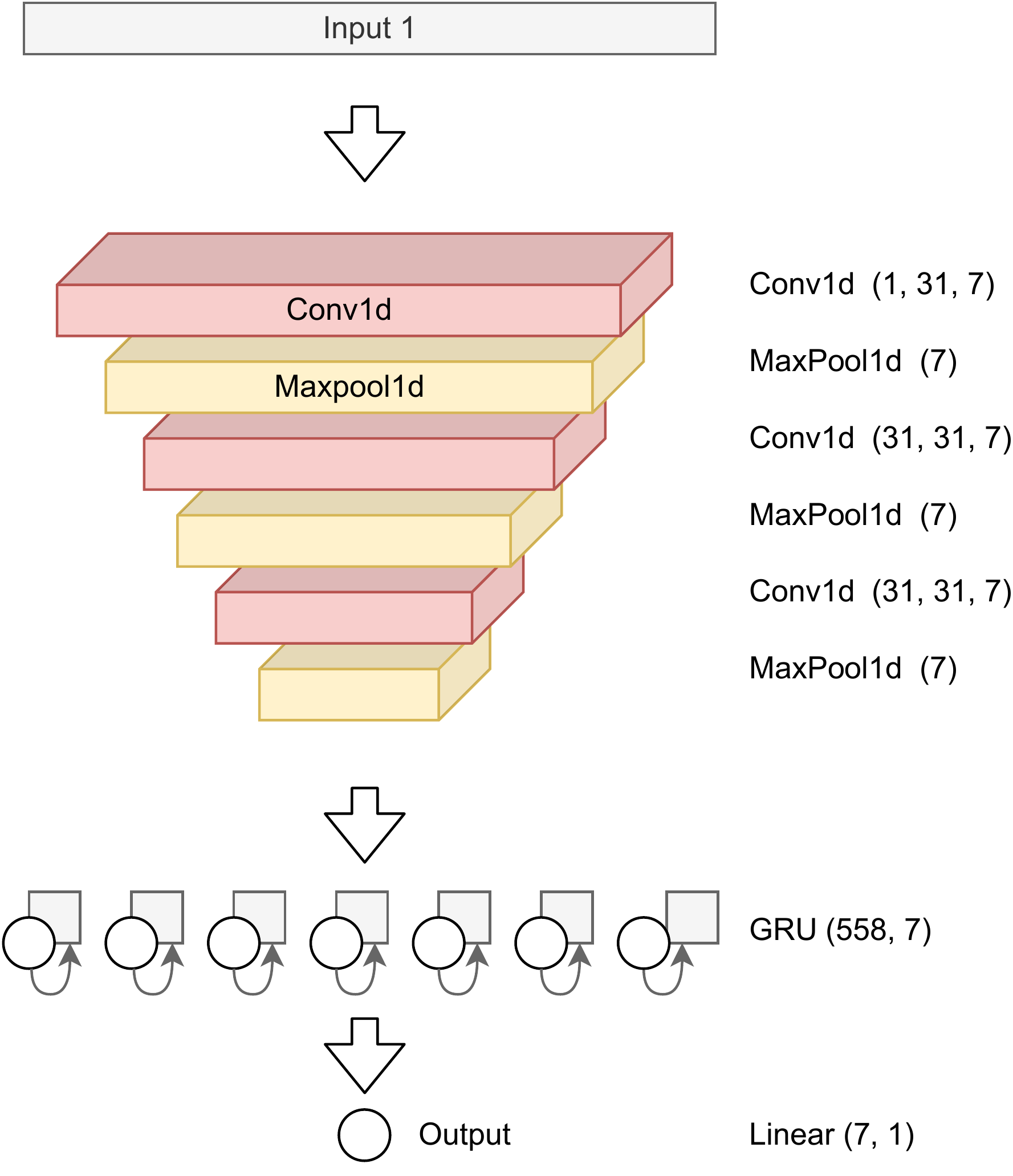}
  \caption{\textbf{Final single-input ANN architecture.} The dimensions of each
    layer are provided in parenthesis using the PyTorch nomenclature.}
\label{fig:1input_nn_arch}
\end{figure}

The selected architecture is described in Fig~\ref{fig:1input_nn_arch}, and its
detection performance is summarized in Table~\ref{tab:results}, row (6).
Compared to our 2-input model, the single-input model exhibits the same
performance, with even a marginal improvement on phase~2, while executing twice
as fast (20~ms versus 40~ms). The detailed hyperparameters of this model are provided in Supplementary Information.

To verify that the use of virtual parallelization via time-dilation (c.f. Supplementary Information) is indeed
necessary to obtain our results, we shrink the time-dilation (set to $168$~ms by
PMBO) to the minimum, \ie $20$~ms since this is the execution duration of the
ANN per forward pass. This removes the virtual parallelization, since the same
ANN must now be used for each sample. Therefore, each step of back-propagation
reaches 8 times less far back in time during training. The result of this
ablation is presented in Table~\ref{tab:results}, row (7). The highly
deteriorated results illustrate the importance of time-dilation. This hints at
the relevance of looking relatively far back in time to annotate sleep spindles.

Finally, to ensure the generality of our ANN, and knowing that spindles change
in older adults~\cite{fernandez2020sleep}, we compare the results using the data
of MODA phase~1 (younger subjects) and the data of MODA phase~2 (older
subjects). Namely, we either train the model on subjects drawn only from
phase~1, or subjects drawn only from phase~2. The results of these experiments
are presented in Table~\ref{tab:results}, rows~(8, 9). We observe that the ANN
trained on phase~1 performs almost as well as the ANN trained on the whole
cohort (6) on all subsets, including phase~2, whereas the ANN trained on phase~2
is noticeably worse on all subsets, even including phase~2. We hypothesize that
this is because phase~2 is comprised of older adults, who have lower amplitude and fewer sleep spindles. Using phase~2 during training is still useful in terms of
generalization. Indeed, the ANN trained on phase~1 only (8) has a slightly worse
performance when tested on phase~1 than the ANN trained on the whole cohort (6).

Note that all models presented beforehand are classifiers. We also train a
regressor with the same architecture, as explained in
Section~\ref{sec:training}. There is a subtle difference in what this model
measures when compared to our classifiers: whereas classifiers predict whether
the signal is a sleep spindle according to the full definition given by MODA
(including post-processing), the regressor predicts the mean score given by the
experts (excluding post-processing). Since we are primarily interested in
classification in this article, we find the threshold that maximizes the
f1-score on the binary labels, presented in Supplementary Information. We find that the optimal
threshold is $0.27$ for phase~1, $0.23$ for phase~2 and $0.26$ for the whole
cohort. We then evaluate the regressor with these thresholds on the
classification task and report the results in Table~\ref{tab:results}, row (10).
These results are slightly weaker than those of the classifier (6). We surmise
that this effect comes from the post-processing steps performed by MODA to
compute the binary labels. We choose the 1-input classifier (6) for the
remainder of this article.

\subsection{Real-time stimulation}
\label{sec:real-time-stimulation}

The performance measured in the previous section is not entirely
representative of the performance on the final task. So far, we have only
measured the capability of the model to annotate each data point of the signal
individually. Yet, we want the ability to send one single stimulation per sleep
spindle.

The ANN delays must be compounded with the other sources of delays (here
reported for the FPGA version as a worst case, as they are slightly lower for
the Coral version), \ie the software delay from FIRs (40~ms), the ANN forward
pass duration (20~ms) and the stimulation hardware delay, to measure our real
stimulation performance. We measure an auditory stimulation delay of 4~ms when
using a basic sound controller, for a total constant delay of 64~ms. The
measured delays are summarized in Table~\ref{tab:delays}, were one can see that
the most significant source of delay is the detection delay of our ANN. Training
a faster model is thus a potential avenue for future work.

\begin{table}
\caption{Delays measured in the Portiloop (sleep spindle configuration)}
\label{tab:delays}
\centering
\begin{tabular}{|l|c|c|}
\hline
\rowcolor[HTML]{CCCCCC} 
\multicolumn{1}{|c|}{\cellcolor[HTML]{CCCCCC}Component} & Hardware delay & Software delay \\ \hline
Electrodes + ADC & - & - \\ \hline
Filters & - & 40~ms \\ \hline
ANN & 20~ms & $\sim$ 250 ($\pm$ 100)~ms \\ \hline
Stimulus & 4~ms & - \\ \hline
Total: $\sim$ 314~ms & 24~ms & 290 ($\pm$ 100)~ms \\ \hline
\end{tabular}
\end{table}

From now on, we redefine:
\begin{inparaenum}[(a)]
\item True positive: the first stimulus sent within the duration of a spindle, taking all delays into account;
\item False positive: any other stimulus;
\item False negative: any spindle that does not receive a stimulus within its
  labeled duration.
\end{inparaenum}

\begin{figure}
  \centering
  \includegraphics[trim=0cm 0cm 0cm 0cm, clip, width=1.0\linewidth]{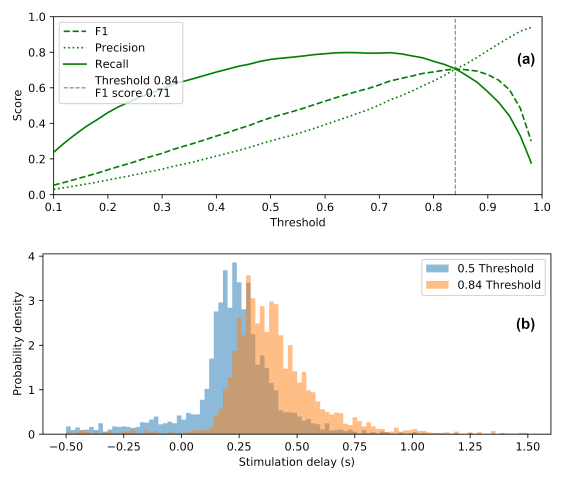}
  \caption{\textbf{Detection threshold trade-off.} (a) Evolution of the stimulation performance with respect to the chosen detection threshold on the ANN output value. A threshold of 0.84 yields the optimal trade-off; however, researchers may wish to select different parameters according to experimental objectives. (b) Distribution of stimulation delays for a classifier with 0.5 and 0.84 thresholds, respectively. Increasing the threshold yields longer delays. Note that delays are negative when spindles are stimulated in advance of human expert annotation.}
\label{fig:threshold_tuning}
\end{figure}

Fig~\ref{fig:threshold_tuning}~(a) displays
the detection performance (taking all delays into account) of our final device.
We compute the stimulation precision, recall and f1-score according to the
aforementioned definitions of true positives, false positives and false
negatives. This provides a visualization of possible trade-offs in terms of how
many spindles we want to stimulate (recall) versus how sure we want to be that
all stimuli are relevant (precision). In terms of f1-score, the best such
trade-off is attained at a threshold of $0.84$ with our model, yielding a
precision and a recall of $0.71$ both.

The timing performance of our system can be observed in
Fig~\ref{fig:threshold_tuning}~(b), which displays the
distribution of stimulation delays, \ie the distribution of the stimulus being
closest to the beginning of each sleep spindle, all delays being taken into
account. Some stimulation delays are negative, as spindles are sometimes
stimulated in advance (note that we count these as false positives, which slightly
harms our reported results).
Fig~\ref{fig:threshold_tuning}~(b) shows the effect of
increasing the detection threshold of our model on the stimulation delays.
According to
Fig~\ref{fig:threshold_tuning}~(a),
choosing a $0.84$ detection threshold over the $0.5$ default classification
threshold in our ANN yields a better stimulation f1-score and in particular much
more precise stimuli, but this comes at the price of slightly shifting the
stimulation delay distribution to the right, \ie introducing some additional
delay to the stimulation, as further seen in Supplementary Information.

To further illustrate the final performance of the system, Figure~\ref{fig:stimu_84} displays an example of its real-time stimulation capability on actual EEG signal (test dataset). More examples and visual insights are provided in Supplementary Information.

Finally, we estimate the Portiloop energy efficiency by running the FPGA
version continuously, powered by a fully-charged 20000~mAh battery. The battery
dies out after 26~hours and 22~minutes, suggesting that our power consumption is
roughly 756~mA. The Coral version runs for approximately 8 hours with a
12000~mAh battery pack suggesting a 1500~mA current draw.

\section{Discussion and future work}

In this article, we introduce the Portiloop, a device that enables the real-time
detection and stimulation of patterns of interest in electroencephalography signals. Our system is
open-source, portable, low-cost, and can be tailored for many brain stimulation research applications.
We propose a pipeline to design neural architectures that are relevant for
processing EEG signals in real time. We further propose an algorithm that
automates the process of finding efficient models (i.e., PMBO), using one-to-many
parallel workers. We demonstrate our proposed system on the closed-loop
stimulation of sleep spindles, a difficult task of high relevance for the
neuroscience community. Our resulting system is the first portable device to be
able to detect and stimulate sleep spindles in real time with an f1-score of
$0.71$, measured on MODA, a dataset renowned for the reliability of its labels.
The Portiloop system can be adapted to any application of EEG closed-loop
stimulation, and potentially, any other neurophysiological signal.
As opposed to classical heuristics, our deep learning-based approach does not
require specific knowledge of the phenomenon of interest when defining the
classifier, nor does it require a way to extract the relevant information. Instead, a
large dataset of annotated signals suffices to derive a high-performance model
that detects complex patterns such as sleep spindles. 

Although we compare our architecture to a state-of-the-art sleep spindle detector (SpindleNet), we did not have access to their weights and thus we could not compare their original model with ours directly on the MODA dataset.
Instead, we retrained their architecture from scratch on MODA, using our own pipeline.
Contrary to the observations of Kulkarni~\etal~\cite{kulkarni2019deep}, we were able remove the envelope
and power inputs without harming the performance of our models. 

Concerning PMBO, the algorithm produces high-performance lightweight
architectures, but we note that the predictions of the meta-learner are often
near-constant in well-performing areas of the search space, suggesting that the
meta-model could not further predict the software cost. In other words, we were unable to differentiate between the best-performing configurations of the neural network. We surmise that this is
due to the large variance in model performance from one training session to
another. This might be further improved by additional training.
In future work, techniques such as Integrated Gradients~\cite{sundararajan2017axiomatic} 
could be used to better understand the search space, and potentially fine-tune the ANN.

Explainable artificial intelligence techniques such as this may also help
researchers to reveal unknown dependencies in neural activity, for example that a spindle might be preceded by another pattern of neural activity (see Supplementary Information for an exploration of which parts of the signal are used by the neural network for classification).

In addition, while the MODA dataset provides high-quality labels, training on a larger
dataset of similar quality would likely further improve the performance of our
models. Expanding MODA is a relevant avenue for future work, as is implementing transfer learning techniques (i.e., tools that allow a trained network to adapt to a different environment), because the EEG acquisition and signal may differ somewhat
from the training data or between individuals. Transfer can be achieved with techniques such as domain
randomization~\cite{tobin2017domain}. Alternatively, a dataset can be collected
on the Portiloop and annotated following the same protocol as MODA.

Long term, we intend to target specific portions of sleep spindles for
stimulation (e.g., beginning, middle, end; or by oscillatory phase). This harder task will likely involve labeling these portions and
developing more advanced RNNs/Transformers so as to consistently predict sleep
spindles. Although our model does use information far
back in time to make predictions, we believe that the main role currently played
by the RNN is to accumulate information regarding whether the last few windows
were spindles or not, rather than actually predicting the future (see Supplementary Information). Such models will likely be more complex and computationally hungry,
which is why the newer hardware implementation of the Portiloop integrates an
embedded tensor processing unit (a powerful neural network accelerator). In general, finding an optimal model for a given Portiloop
application involves either retraining our ANN, or re-executing PMBO to find a
whole new architecture. Both activities can be done by interested practitioners using tools that accompany this work.

In sum, we hope that the Portiloop will help the neuroscience community
explore brain functions, such as the role of sleep spindles in memory consolidation.

\section*{Acknowledgment}
We thank Karine Lacourse for expert advice on spindle detection, and the MODA
team for database access.
Fig~\ref{fig:hardware}
uses icons from \url{https://flaticon.com}.


\bibliography{Portiloop}

\onecolumn

\appendices

\section{Model-based explanation of sleep spindles}
\label{app:grad_explain}

\begin{figure}[H]
      \centering
      \includegraphics[trim=0 0 0 0, width=1.0\linewidth]{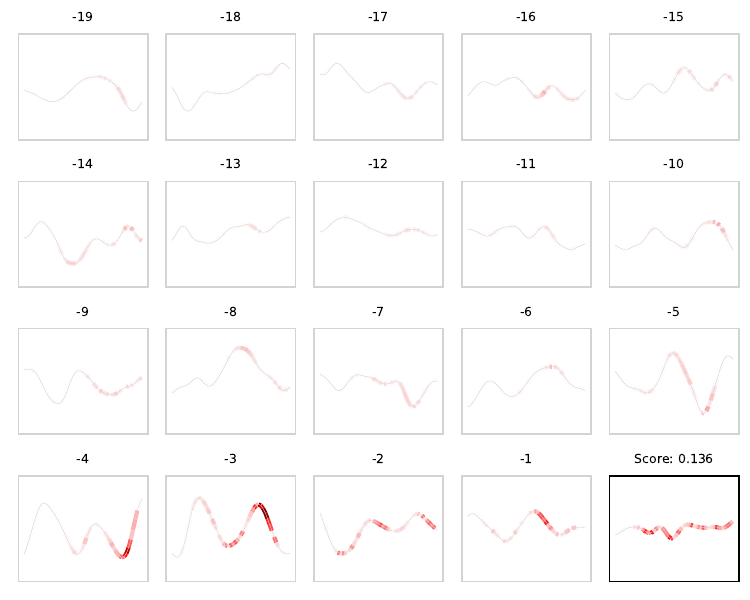}
      \caption{Integrated gradients (classifier ANN output: $0.136$). The \emph{integrated gradients} algorithm enables exploring why the model takes a given decision (the more a portion of the signal is represented in red, the highest its influence on the current output of the ANN).
      Grey windows are past inputs, whereas the black window is the current input: the past influences the current output due to the RNN.
      Here, the model finds that it is looking at the aftermath of a sleep spindle.
      With our time dilation and window size, a small portion of the window overlaps from one sample to the next. We see that this portion (at the left-hand side of each window) is in fact ignored by the model. Therefore, it is probably possible to shrink our model even more, although PMBO did not find this.
      In the future, this type of visualizations might also help experts better understand what sleep spindles are by revealing unknown influences.}
    \label{fig:grad_explained_tn}
\end{figure}

\begin{figure}[H]
      \centering
      \includegraphics[trim=0 0 0 0, width=1.0\linewidth]{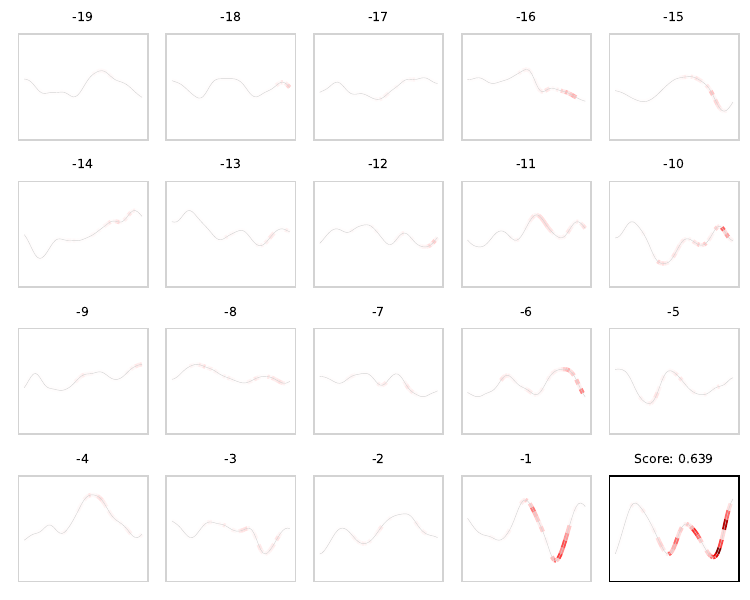}
      \caption{Integrated gradients (classifier ANN output: $0.639$). The current window is within an actual sleep spindle. The model mainly focuses on the spindle itself, but also a few events that happened further back in time, to make its decision.}
    \label{fig:grad_explained_tp}
\end{figure}

\newpage
\section{Stimulation visualizations}
\label{app:signal_visu}

\begin{figure}[H]
  \centering
  \includegraphics[trim=0 0 0 0, width=0.9\linewidth]{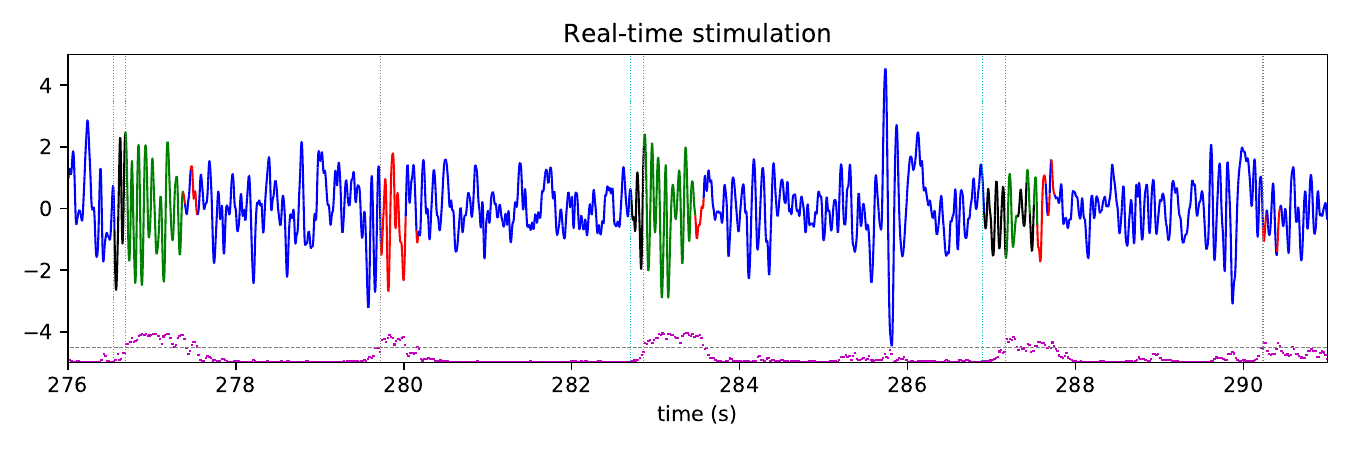}
  \caption{Classifier with threshold 0.5, overall stimulation performance. Blue: no sleep spindle and no detection. Black: sleep spindle not detected. Green: sleep spindle correctly detected. Red: detection where the signal is not a spindle. Vertical cyan: beginning of a spindle. Vertical grey: stimulus. Magenta: ANN output. Horizontal grey: detection threshold. }
\label{fig:class_stim_50}
\end{figure}

\begin{figure}[H]
  \centering
  \includegraphics[trim=0 0 0 0, width=0.9\linewidth]{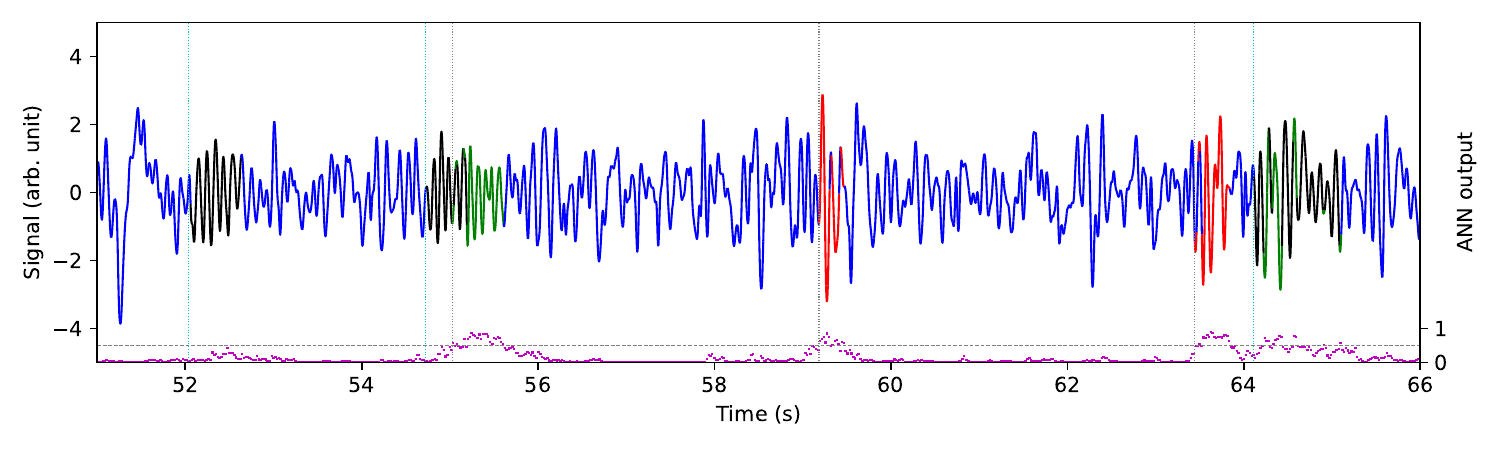}
  \caption{Different stimulation failure modes (classifier with threshold 0.5). This figure illustrates typical `failure' cases of our final sleep spindle stimulating device. False negative: the first spindle is missed because the threshold is too high. True positive: the second spindle is correctly stimulated. False positive: a part of the signal not labeled as a sleep spindle by MODA is detected as a spindle by the device and stimulated. Almost true positive: the sleep spindle is stimulated in advance (NB: we count this case as a false positive).}
\label{fig:class_stim_5_fail}
\end{figure}


\begin{figure}[H]
  \centering
  \includegraphics[trim=0 0 0 0, width=0.9\linewidth]{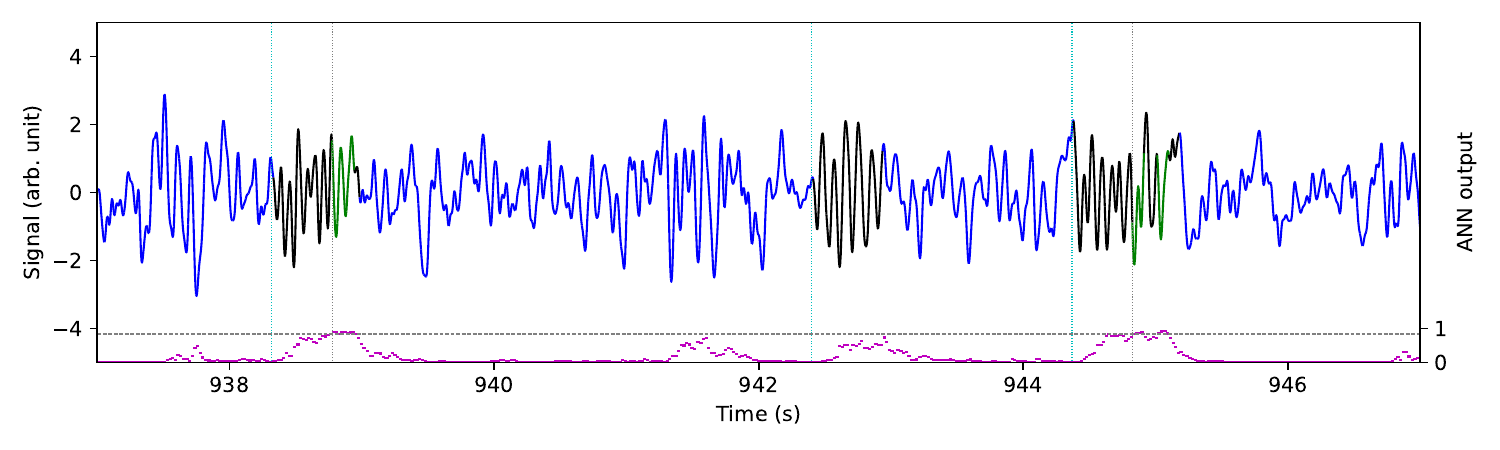}
  \caption{Classifier with threshold 0.84, failure example. Increasing the detection threshold comes with more false negative stimuli (vertical blue not followed by vertical grey).}
\label{fig:class_stim_84_fail}
\end{figure}

\section{PMBO hyperparameters}
\label{app:hyperparams_pmbo}
\begin{table}[H]
\caption{Hyperparameters used for PMBO}
\label{tab:hyperparams_pmbo}
\centering
\begin{tabular}{|l|c|}
\hline
\rowcolor[HTML]{CCCCCC} 
\multicolumn{1}{|c|}{\cellcolor[HTML]{CCCCCC}hyperparameter} & selected value \\ \hline
range cost hardware & 1000-80000 \\ \hline
noise type 1 & 0,25 \\ \hline
noise type 2 & 0,1 \\ \hline
m & 200 \\ \hline
meta network type & MLP \\ \hline
\# layers meta network & 3 \\ \hline
hidden size meta network & 200 \\ \hline
optimizer meta network & SGD \\ \hline
learning rate meta network & 0.05 \\ \hline
weight decay meta network & 0.01 \\ \hline
\end{tabular}
\end{table}

Two types of noise are used in our implementation of PMBO to foster exploration of the hyperparameter space:
\begin{itemize}
    \item noise type 1: portion of the m sampled models that are sampled randomly in the whole hyperparameter space, instead of in a Gaussian around the last completed experiment.
    \item noise type 2: portion of the time when a model is sampled randomly in the m models, instead of being selected by its Pareto efficiency.
\end{itemize}
The ANN used for our meta model is a simple Multi-Layer Perceptron (MLP) of 3 fully connected layers.
The hyperparameters we use in PMBO are summarized in Table~\ref{tab:hyperparams_pmbo}.

\section{Model hyperparameters}
\label{app:hyperparams_model}
\begin{table}[H]
\caption{Hyperparameters used to train the final model}
\label{tab:hyperparams_model}
\centering
\begin{tabular}{|l|c|c|}
\hline
\rowcolor[HTML]{CCCCCC} 
\multicolumn{1}{|c|}{\cellcolor[HTML]{CCCCCC}Hyperparameter} & Selected value & PMBO \\ \hline
\rowcolor[HTML]{E5E5E5} 
\multicolumn{3}{|c|}{\cellcolor[HTML]{E5E5E5}{\color[HTML]{333333} Training}} \\ \hline
optimizer & AdamW &  \\ \hline
\# epochs max & 150 &  \\ \hline
epochs before early stopping & 20 &  \\ \hline
early stopping running average factor & 0,1 &  \\ \hline
batches per epoch & 1000 &  \\ \hline
batch size & 256 & X \\ \hline
dropout on first layer & False &  \\ \hline
dropout factor & 0,5 &  \\ \hline
adam learning rate & 0,0005 & X \\ \hline
adam weight decay & 0,01 &  \\ \hline
balancing mode & oversampling &  \\ \hline
type of training & classification &  \\ \hline
sequence length & 50 &  \\ \hline
\rowcolor[HTML]{E5E5E5} 
\multicolumn{3}{|c|}{\cellcolor[HTML]{E5E5E5}{\color[HTML]{333333} Architecture}} \\ \hline
window size (s) & 0,216 & X \\ \hline
time dilation (s) & 0,168 & X \\ \hline
\# CNN layers & 3 & X \\ \hline
\# CNN channels & 31 & X \\ \hline
stride convolution & 1 & X \\ \hline
kernel size convolution & 7 & X \\ \hline
dilation convolution & 1 & X \\ \hline
stride max pooling & 1 & X \\ \hline
kernel size max pooling & 1 & X \\ \hline
dilation max pooling & 1 & X \\ \hline
RNN layers & 1 & X \\ \hline
RNN hidden size & 7 & X \\ \hline
\end{tabular}
\end{table}
The hyperparameters we use in our final model are listed in Table~\ref{tab:hyperparams_model}.
Hyperparameters that were chosen by PMBO are marked with a cross under the PMBO column.

\newpage
\section{Best threshold for classifiers and regressors}

\label{app:threshold_detection}
\begin{figure}[H]
  \centering
  \includegraphics[trim=0 0 0 0, width=0.7\linewidth]{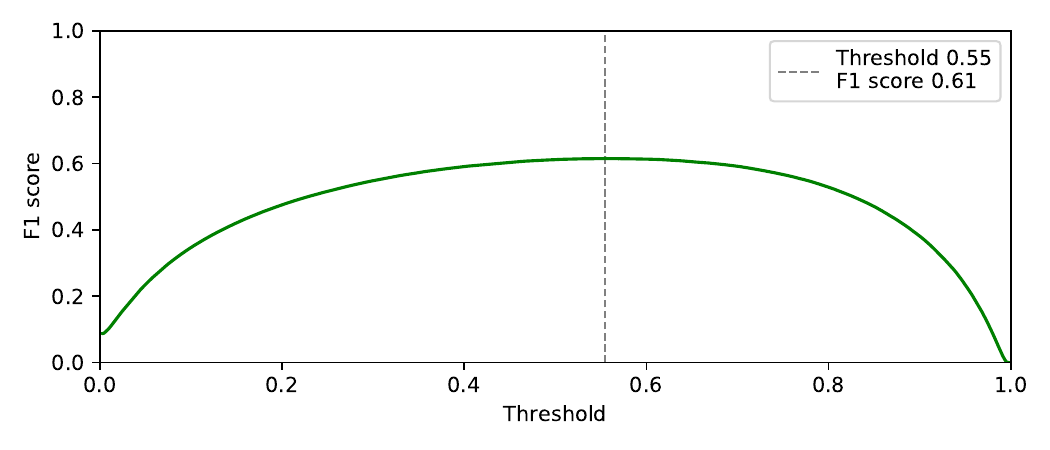}
  \caption{F1 score evolution with threshold variation on classification}
\label{fig:f1scorevsthreshold_classification_full}
\end{figure}

\begin{figure}[H]
  \centering
  \includegraphics[trim=0 0 0 0, width=0.7\linewidth]{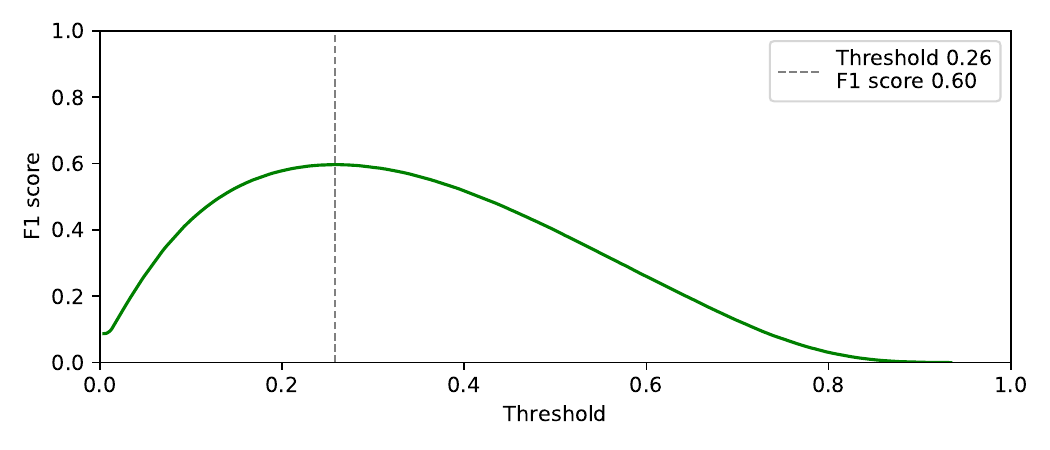}
  \caption{F1 score evolution with threshold variation on regression}
\label{fig:f1scorevsthreshold_regression_full}
\end{figure}

\section{Detection delay distributions}

\begin{figure}[H]
      \centering
      \includegraphics[trim=0 0 0 0, width=0.9\linewidth]{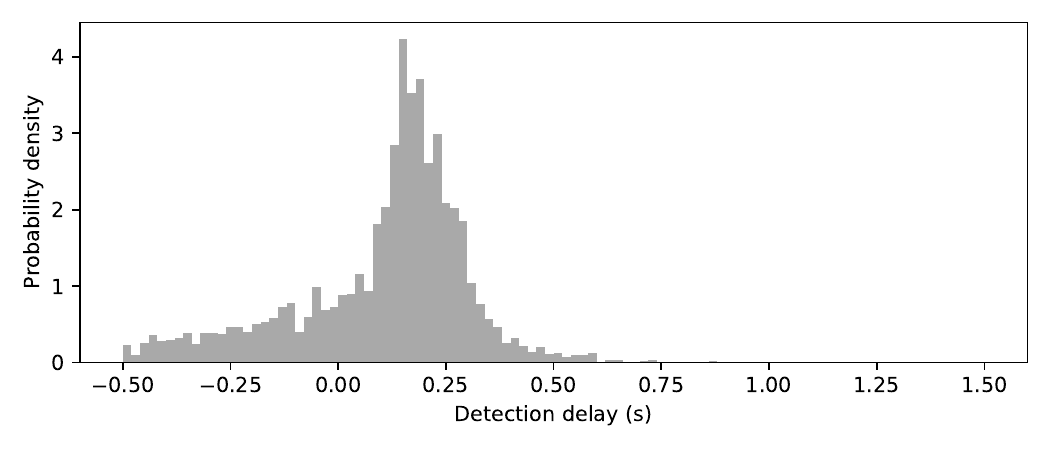}
      \caption{Stimulation histogram for 1 input network classification with a 0.25 threshold}
    \label{fig:stimulation_histogram_1_input_full_0.25}
\end{figure}

\section{2-input network architecture}
\label{app:2-input-arch}
\begin{figure}[H]
  \centering
  \includegraphics[trim=0 0 0 0, width=0.5\linewidth]{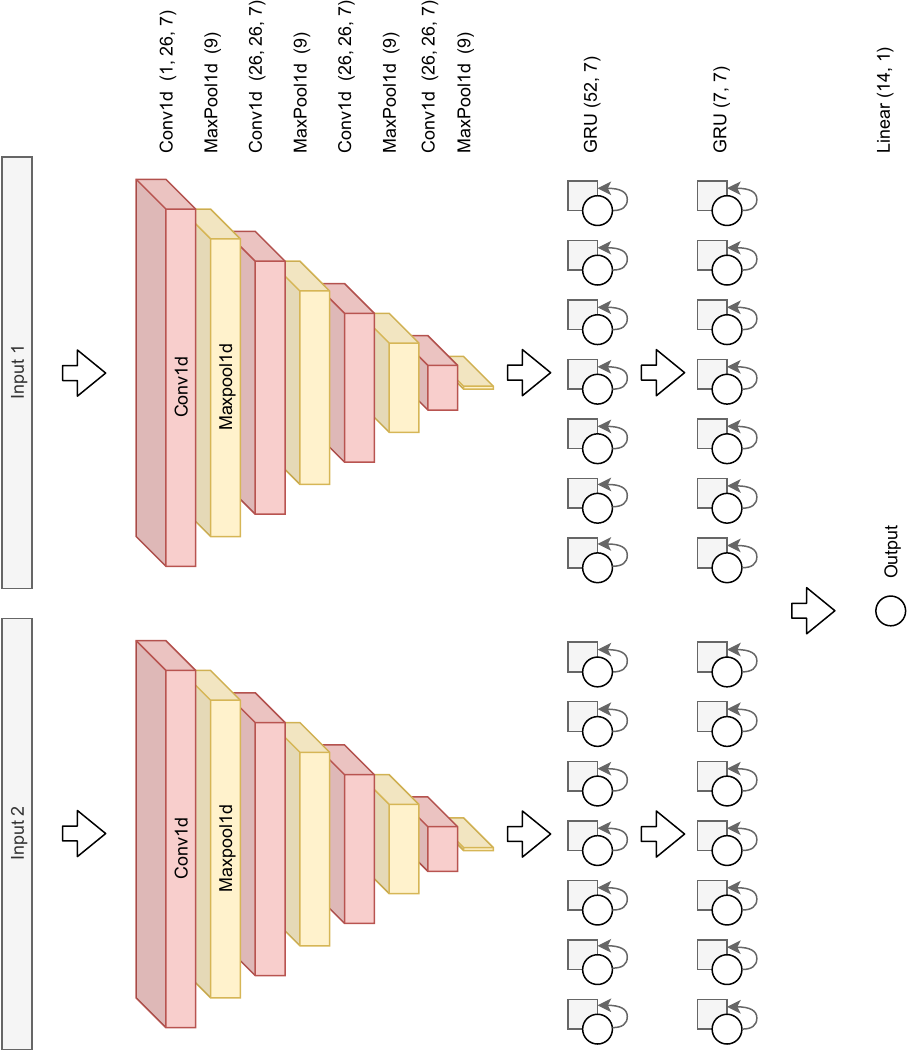}
  \caption{2-input neural network architecture. This architecture consists of two sequences of CNN and RNN. The first sequence processes the cleaned raw signal (input 1), whereas the second processes the envelope (input 2). The latent features from both branches are concatenated and fed to a fully connected layer to yield the output of the ANN.}
\label{fig:2inputs_nn_arch}
\end{figure}

\section{Threshold tuning effect on stimulation}

\begin{figure}[H]
  \centering
  \includegraphics[trim=0 0.4cm 0 0.3cm, clip, width=0.7\linewidth]{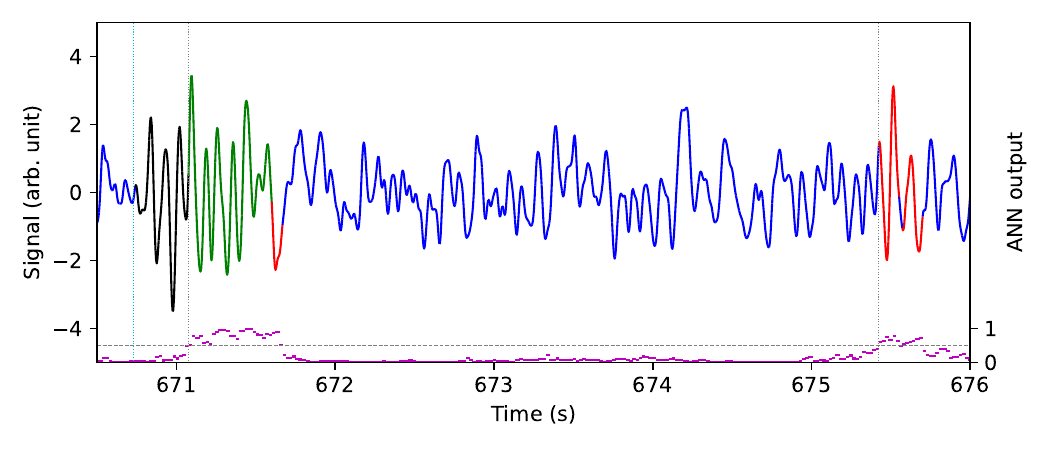}
  \caption{\textbf{Stimulation with a 0.5 threshold.} The first part of the spindle is not detected (false negative), which introduces an ANN software delay. In addition, a small
    portion of the signal after the spindle is still detected as such (false
    positive), but it has no impact on our stimulation procedure. Finally,
    another portion of the signal that was not annotated as a spindle by MODA is
    detected as a spindle by our model (false positive).}
\end{figure}

\begin{figure}[H]
  \centering
  \includegraphics[trim=0 0.4cm 0 0.3cm, clip, width=0.7\linewidth]{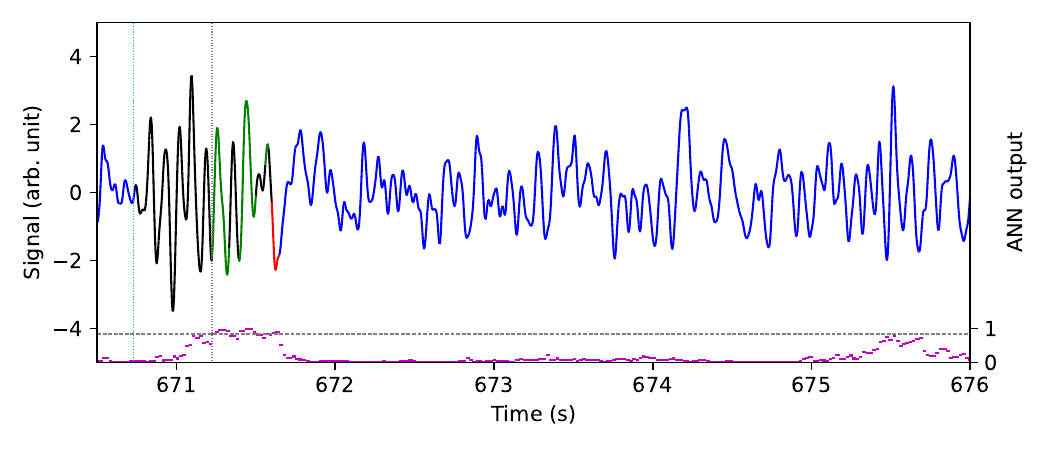}
  \caption{\textbf{Stimulation with a 0.84 threshold.} Increasing the threshold removes false positives at the price of some stimulation delay.}
\end{figure}

\section{Online standardization through exponential moving average}

We standardize the
signal on the fly though exponential moving average.  In other words,
we transform the filtered signal $s(t)$ to $s'(t)$ according to:
\begin{equation}
    s'(t) = \frac{s(t) - \hat{\mu}(t)}{\hat{\sigma}(t)}
    \label{eq:standardization}
\end{equation}
where $\hat{\mu}$ and $\hat{\sigma}$ are estimates of the mean and
standard deviation of the filtered signal, computed as follows:
\begin{equation}
    \delta(t) = s(t) - \hat{\mu}(t-1)
\end{equation}
\begin{equation}
    \hat{\mu}(t) = \hat{\mu}(t-1) + \alpha_{\mu}\delta(t)
    \label{eq:running_av}
\end{equation}
\begin{equation}
    \hat{\sigma}^2(t) = (1 - \alpha_{\sigma})(\hat{\sigma}^2(t-1) + \alpha_{\sigma}\delta^2(t))
\end{equation}
\begin{equation}
    \hat{\sigma}(t) = \sqrt{\hat{\sigma}^2(t)}
\end{equation}
$\alpha_\mu$ and $\alpha_\sigma$ being hyperparameters in $[0,1]$.
This custom real-time standardization makes the signal comparable from
one subject to another, enabling generalizable learning.

\section{Parallel Model-Based Optimization}

When developing a novel Portiloop application, the practitioner
needs to devise a neural network that is both high-performance and lightweight,
by selecting the right set of hyperparameters $H$ in the space of all possible
hyperparameter sets $\mathcal{H}$. Such hyperparameters include the size of the
sliding window, the number of layers in each part of the ANN, the width of each
layer, the time dilation (see Section~\ref{sec:dilation}), the type of
optimizer, the hyperparameters of the optimizer itself, etc. $\mathcal{H}$ can
be very large, and finding a set of hyperparameters that yields a
high-performance model within given hardware constraints is far from trivial. We
introduce ``Parallel Model-Based Optimization'' (PMBO), a network-based
algorithm that automates this process in a parallel fashion. Released as
open-source along with our code, PMBO is essentially a parallelized and evolved
version of ``Probabilistic SMBO''~\cite{yin2020probabilistic}. PMBO is a
guided-search approach that finds a suitable set of hyperparameters rapidly. For
this matter, it uses one machine (or process) to train a \emph{meta network}
whose role is to predict non-trivially available costs for any given set of
hyperparameters.
Furthermore, PMBO uses any available machines (or processes) in parallel to
train ANNs from sets of hyperparameters selected based on the cost estimated by
the meta model.

The purpose of PMBO is to find Pareto-optimal sets of hyperparameters that
minimize both a~\emph{software cost} and a~\emph{hardware cost}. Hyperparameter
selection is a bi-objective problem in our setting: on one hand, we want an ANN that performs well at detecting the
desired patterns. We measure this performance in terms of the f1-score of our
model. The f1-score depends both on the precision (how sure we are that positive
outputs are true positives) and on the recall (how sure we are that we capture
all true positives) of the model. It is defined as:
\begin{equation}
    \text{f1} = \frac{2*\text{recall}*\text{precision}}{\text{recall}+\text{precision}}
\label{eq:f1}
\end{equation}
where
\begin{equation}
    \text{precision} = \frac{\text{true positives}}{\text{true positives}+\text{false positives}}
\end{equation}
and
\begin{equation}
    \text{recall} = \frac{\text{true positives}}{\text{true positives}+\text{false negatives}}
\end{equation}
The f1-score is in $[0,1]$, with $1$ being a perfect classifier.
We cast this into a minimization problem by defining our \emph{software cost} as $L_s= 1-\text{f1}$.

On the other hand, we want our ANN to be as lightweight as possible,
so it fits in the limited memory of the Portiloop and executes as quickly as
possible. A precise measurement of the execution duration and amount of
memory needed on the board is difficult. In fact, we do not have access to
these results until the model is actually synthesized on the board, which is a
lengthy process. Thus, we use the number of trainable weights in the ANN as a proxy
for these concerns, and call this number our \emph{hardware cost}~$L_h$.
Note that our choice of costs is arbitrary and other custom costs can be used
instead.

Let us consider a meta dataset $\mathcal{E}$ of previously completed
experiments (\ie tuples $E=(H,L_s,L_h)$ of hyperparameter sets with
their real costs).  Let us also consider the following
Pareto-dominant relation:
\begin{definition}
  Let $E=(H,L_s,L_h)$ and $E'=(H',L_s',L_h')$ be two experiments in
  $\mathcal{E}$.  We say that $E$ Pareto-dominates $E'$ and we denote
  $E < E'$ when $ L_s < L_s' \wedge L_h < L_h'$.
(NB: in the context of this minimization problem, dominating means
having the smallest costs, hence the notation)
\end{definition}
\noindent
For a given experiment $E$ and meta dataset $\mathcal{E}$, we denote
$D_\mathcal{E}(E)$ as the number or experiments in $\mathcal{E}$ that
are Pareto-dominated by $E$, and $d_\mathcal{E}(E)$ as the number of
experiments in $\mathcal{E}$ that Pareto-dominate $E$.  In other
words:
\begin{equation}
    D_\mathcal{E}(E) = |\{E^i\in\mathcal{E}, E^i > E\}|
\end{equation}
\begin{equation}
    d_\mathcal{E}(E) = |\{E^i\in\mathcal{E}, E^i < E\}|
\end{equation}
For a given set of hyperparameters $H \in \mathcal{H}$, we further
define an estimate of the corresponding completed experiment $E$ as
$\hat{E}~=~(H, \hat{L}_s, \hat{L}_h)$, where $\hat{L}_s$ and
$\hat{L}_h$ are estimates of the real costs, computed by the meta
network.  Note that in our setting, $\hat{L}_h = L_h$ is available and
thus only $\hat{L}_s$ is estimated by the meta network.  We propose
the heuristic Pareto efficiency $\eta(\hat{E})$:
\begin{equation}
    \eta(\hat{E}) = a_\mathcal{E}(\hat{E}) + b_\mathcal{E}(\hat{E}) +  s_\mathcal{E}(\hat{E}) - h_\mathcal{E}(\hat{E})
\end{equation}
where $a_\mathcal{E}(\hat{E})$ promotes hyperparameter sets whose predicted costs are not dominated by many experiments in $\mathcal{E}$:
\begin{equation}
    a_\mathcal{E}(\hat{E}) = 1 - \frac{d_\mathcal{E}(\hat{E})}{|\mathcal{E}|}
\end{equation}
$b_\mathcal{E}(\hat{E})$
promotes hyperparameter sets whose predicted costs dominate many experiments in $\mathcal{E}$:
\begin{equation}
    b_\mathcal{E}(\hat{E}) = \frac{D_\mathcal{E}(\hat{E})}{|\mathcal{E}|}
\end{equation}
$s_\mathcal{E}(\hat{E})$
promotes hyperparameter sets whose predicted software costs are better than the best software cost amongst all completed experiments in $\mathcal{E}$:
\begin{equation}
     s_\mathcal{E}(\hat{E}) = \frac{\text{min}(\{L_s, (H, L_s, L_h) \in\mathcal{E}\})}{\hat{L}_s}
\end{equation}
and $h_\mathcal{E}(\hat{E})$ penalizes hyperparameter sets that have a high density with respect to experiments present in $\mathcal{E}$ in terms of their hardware cost.
More precisely, we define a range of hardware costs we are interested in, and we split this range into a number of bins.
We then compute the binned density of experiments in $\mathcal{E}$ over this range, and multiply this density by the range's width.
The penalty $h_\mathcal{E}(\hat{E})$ is the height of the resulting bin where $\hat{L}_h$ stands.
Multiplying the density by the range's width enforces $h_\mathcal{E}(\hat{E}) > 1$ in regions of high density and $h_\mathcal{E}(\hat{E}) < 1$ in regions of low density.

\begin{figure}[H]
  \centering
  \includegraphics[trim=0 0 0 0, width=0.8\linewidth]{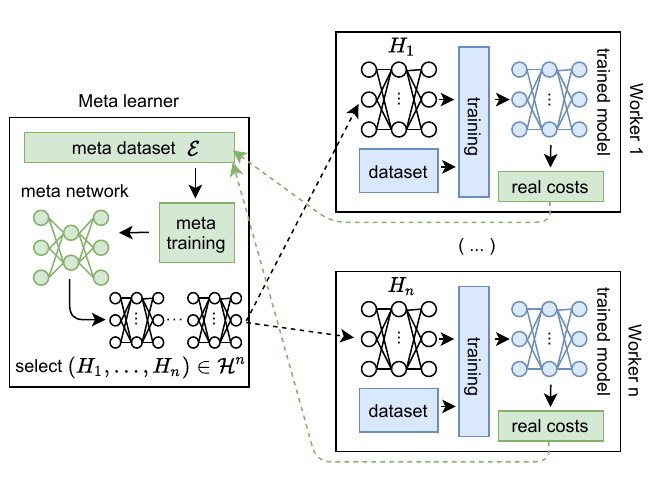}
  \caption{\textbf{The Parallel Model-Based Optimization (PMBO) algorithm.} The algorithm serves to automate the selection of hyperparameters within given hardware constraints.
  It is based on a standard single-producer and
    multiple-consumers architecture. A ``meta learner'' is in charge of
    producing relevant hyperparameter sets in a guided fashion. It sends these sets to ``workers'', and keeps producing new sets as
    long as idle workers remain. Each worker that has received a new
    set starts training an ANN corresponding to the assigned set. Once this training
    ends, the real costs of the hyperparameter set can be computed and
    are sent back to the meta learner. The best-performing models can then be selected for implementation.}
\label{fig:pmbo}
\end{figure}

Fig~\ref{fig:pmbo} explains our PMBO algorithm. A central meta learner is
communicating with $n$ peripheral workers to find a hyperparameter set
$H^\ast\in\mathcal{H}$ that is Pareto-optimal for both the software and hardware
costs (\ie non-Pareto-dominated by any other set). For this matter, the
algorithm uses a meta dataset $\mathcal{E}$ of tuples $E^i=(H^i, L_s^i, L_h^i)$
to train a meta network that maps any hyperparameter set $H\in\mathcal{H}$ to
its corresponding (estimated) costs $\hat L_s$ and $\hat L_h$ \footnote{$\hat
  L_h$ is an output of the meta network in the general case. However, with our
  choice of hardware cost it is not, since the ground truth $L_h$ is
  available.}. Once the meta network is trained, it is used to guide the
sampling process. More exactly, we sample $m$ hyperparameter sets in
$\mathcal{H}$ from a multivariate Gaussian distribution around the
hyperparameter set corresponding to the last results received from the workers
by the meta learner. Sets that don't satisfy user-defined constraints (\eg that
have already been tested, or, when $L_h$ is available, that fall outside the
range we are interested in) are discarded and resampled. We then select the best
set in terms of Pareto efficiency, estimated thanks to the trained meta network.
This selected set is appended to a buffer, waiting to be consumed by an
idle worker. The sampling process is repeated until the buffer is full.
When a worker is idle, it fetches an ANN architecture and training instructions
from the buffer. The worker yields a measurement of the real hardware
and software costs for the current hyperparameter set, which are sent back to
the meta learner and appended to the meta dataset. The meta learner then uses
the updated meta dataset to train a new meta network, and so on.

\section{Virtual ANN parallelization with time dilation}
\label{sec:dilation}

\begin{figure}[H]
  \centering
  \includegraphics[trim=0 0cm 0cm 0cm, clip, width=0.7\linewidth]{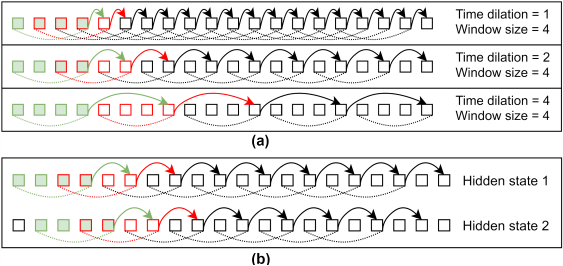}
\caption{\textbf{A lightweight solution to using historical signal information.} To ``look'' farther back in time, which facilitates detection accuracy, we introduce (a) Time dilation, and (b) Virtual parallelization. In (a), a sliding window of the 4
  last samples (dotted curves) is used as input to
  the model. The time dilation (arrows) is the number of samples between two
  forward passes in the ANN. When it is small (top), two consecutive
  windows overlap (see \eg green and red
  windows), meaning consecutive passes contain redundant information. When the time dilation is large
  (bottom), this issue is corrected, and back-propagation will reach
  much further back in time for the same number of forward passes. Virtual parallelization reduces delay. In (b), the time dilation
    is 2, so we keep track of 2 independent hidden states and feed these
    alternately to the ANN. This trick removes the delay of time dilation.}
\label{fig:dilation_parallel}
\end{figure}

Given the Portiloop's design constraints, we sought a lightweight means of allowing our resource-restricted network to use as much signal history as possible (as do larger neural networks).
Time dilation~\cite{chang2017dilated} is a technique that enables
recurrent units such as Gated Recurrent Units (GRUs) to look further back in time before
gradients vanish, at no computational cost. We propose a version of
this technique that allows us to virtually parallelize a single
physical ANN into several decoupled virtual models. Our approach
enables shallow recurrent neural networks to look further back in time
by skipping the redundant information that is inherent to the use of a
sliding window as input, while still acting as fast as possible.
Fig~\ref{fig:dilation_parallel}~(a) illustrates how time dilation can be used to look
further back in time and avoid redundancy.

Although time dilation enables reaching further back in time at no
extra computational cost, this comes with a cost in terms of delays.
Since our technique causes samples
to be skipped between
forward passes in the ANN, a detection delay that can be as long as
the time dilation is introduced.  We correct this issue by
implementing a trick that we call \emph{virtual parallelization}.  We
create a First In First Out (FIFO) list as large as the time dilation,
and fill this list with independent hidden states\footnote{\emph{Hidden states} are memory cells of recurrent units like GRUs.}. At each time-step,
we pop a hidden state from this list, feed it to the recurrent units
of our ANN, perform a forward pass, and append the resulting hidden
state to the list.  Doing this without skipping samples is equivalent
to having several decoupled models running in parallel as illustrated
in Fig~\ref{fig:dilation_parallel}~(b), although one single ANN is physically
used.
This trick allows us to keep acting as fast as possible since it
removes the need for skipping samples, while still reaching far back
in time at no extra computational cost.

\end{document}